\documentclass[a4paper,10pt]{article}
\usepackage[utf8]{inputenc}
\usepackage{amsthm}
\usepackage{amsfonts}
\usepackage{amssymb}	
\usepackage{amsmath}
\usepackage{mathtools}
\usepackage[english]{babel}
\usepackage{color}
\usepackage{slashed}
\usepackage{enumerate}
\usepackage{graphicx}
\usepackage{systeme}
\usepackage{dsfont}
\usepackage{relsize}
\usepackage[margin=0.90in]{geometry}
\usepackage{float}
\usepackage{tikz}
\usepackage{graphicx}
\usepackage{bmpsize}
\usepackage{epstopdf}
\usepackage{bbm}
\usepackage{xcolor,etoolbox}
\usepackage{titling}
\usepackage{authblk}

\usepackage{blindtext,graphicx}
\usepackage[absolute]{textpos}
\usepackage{physics}
\usepackage[hang,flushmargin]{footmisc}

\usepackage{xfrac}
\usepackage{nicefrac}
\usepackage{esvect}
\usepackage{cases}
\usepackage{empheq}
\usepackage{stmaryrd}
\usepackage{cancel}
\usepackage{url}
\usepackage{csquotes}
\usepackage{setspace}
\newtheorem{theorem}{Theorem}[section]
\newtheorem{proposition}[theorem]{Proposition}

\theoremstyle{definition}

\newtheorem{remark}[theorem]{Remark}
\makeatletter
\newcommand{\pushright}[1]{\ifmeasuring@#1\else\omit\hfill$\displaystyle#1$\fi\ignorespaces}
\newcommand{\pushleft}[1]{\ifmeasuring@#1\else\omit$\displaystyle#1$\hfill\fi\ignorespaces}
\makeatother

\newcommand{\id}{{\mathbf{\mathbbm{1}}}}

\newcommand{\dev}{{\rm dev}}
\newcommand{\Sym}{{\rm Sym}}
\newcommand{\sym}{{\rm sym}}
\newcommand{\skw}{{\rm skew}}
\newcommand{\Curl}{{\rm Curl}}
\newcommand{\Div}{{\rm Div}}
\newcommand{\axl}{{\rm axl}}
\newcommand{\anti}{{\rm \textbf{Anti}}}
\newcommand{\so}{\mathfrak{so}}

\def\dd{\displaystyle}

\renewcommand{\skew}{\, \mathrm{skew}\,}

\renewcommand{\skew}{\, \mathrm{skew}}

\def\barr{\begin{array}}
	\usepackage{lscape}

	\def\earr{\end{array}}
\def\bec#1{\begin{equation}\label{#1}}
\def\becn{\begin{equation*}}
\def\endec{\end{equation}}
\def\endecn{\end{equation*}}
\def\dd{\displaystyle}
\def\bfm#1{\mbox{\boldmat}}

\title{\vspace{-2cm} Cosserat micropolar elasticity:  classical Eringen vs.   dislocation form}
\author{Ionel-Dumitrel Ghiba\;\!\thanks{Department of Mathematics, Alexandru Ioan Cuza University of Ia\c si, Blvd. Carol I, no. 11, 700506 Ia\c si, Romania; and Octav Mayer Institute of Mathematics of the Romanian Academy, Ia\c si Branch, 700505 Ia\c si},
	\:\,
Gianluca Rizzi\;\!\thanks{Chair  of Architecture and Civil Engineering, Technische Universit\"{a}t  Dortmund, August-Schmidt-Str. 8, 44227 Dortmund, Germany},
\\
Angela Madeo\;\!\thanks{Head of Chair of Continuum Mechanics, Faculty of Architecture and Civil Engineering, Technische Universit\"{a}t Dortmund,  August-Schmidt-Str. 8, 44227 Dortmund, Germany} 
\:\ and\ 
Patrizio Neff\;\!\thanks{Head of Chair for Nonlinear Analysis and Modelling, Fakultät für Mathematik, Universität   Duisburg-Essen,  Thea-Leymann-Straße 9, 45127 Essen, Germany}
}
\thanksmarkseries{arabic}
\date{\today}
\begin{document}
\maketitle
\begin{abstract}
In this paper we do a comparative presentation of the linear  isotropic Cosserat elastic model from two perspectives: the classical Mindlin-Eringen-Nowacki description in terms of a microrotation vector and a new formulation in terms of a skew-symmetric matrix and a curvature energy   in dislocation form. 
We provide the reader with an alternative representation of the energy for the isotropic Cosserat model to ease the comparison with the relaxed micromorphic model and the  geometrically nonlinear Cosserat elastic model.
\end{abstract}
\begin{footnotesize}
\tableofcontents
\end{footnotesize}
\section{Introduction}

Classical continuum mechanics considers material continua as simple point-continua with points having three displacement-degrees of freedom, and the response of a material to the displacement of its points is characterized by a symmetric Cauchy force-stress tensor presupposing that the transmission of loads through surface elements is uniquely determined by a force vector, neglecting couples. Such a model may be insufficient for the description of certain physical phenomena. Non-classical behaviour due to microstructural effects is observed mostly in regions of high strain gradients, e.g. at notches, holes or cracks.
The Cosserat model is one of the best-known generalized continuum models. In that model, the classical translational degrees of freedom are augmented with a so-called microrotation, providing three additional degrees of freedom. 

The Cosserat model  has emerged from the seminal work of the brothers Francois and Eugene Cosserat at the turn of the last century (``Theorie des corps deformables.'' 1909, Review 1912, english translation by D. Delphenich 2007) \cite{Cosserat09}. They attempted to unify field theories embracing mechanics, optics and electro-dynamics through a common principle of least action (Euclidean action). Their main aim was to produce the correct general form of the energy for the variational problem. Postulating the invariance of energy under Euclidean transformations they were able to derive the equations of balance of forces and balance of angular momentum in a geometrically nonlinear format. However, they never wrote down any constitutive equations and never considered a linearized model as we will do here.

Compared to classical linear elasticity the linear Cosserat model features three additional, independent degrees of freedom, related to the rotation of each particle which need not coincide with the macroscopic rotation of the continuum at the same point. In the simplest isotropic case, one coupling constant, here called Cosserat couple modulus $\mu_{\rm c}> 0$ and three internal length scale parameters need to be determined/measured in addition to the two classical elastic Lam\'e-constants.  In the following, let us concentrate first on the static linear setting with quadratic free energy. In this linear setting, the model is in fact already given by the german scientist W. Voigt in 1887 \cite{Voigt87}. Also P. Duhem in 1893 \cite{duhem1893potentiel} had noticed that various phenomena which seemed incompatible with classical continuum mechanics could be described as effects of direction, and he suggested that materials be visualized as sets of points having vectors attached to them, i.e., oriented or polar media. However, for historical precision, the date of birth of a polar continuum is the year 1686, when Jakob Bernoulli introduced angular momentum as a postulate, independent of balance of momentum, see the work by Truesdell \cite{truesdell1964zusammenfassender}. 

One of the essential features of polar continua is that the force stress tensor is not necessarily symmetric\footnote{But it can still be chosen to be symmetric in the nonlinear Cosserat model by setting the Cosserat couple modulus $\mu_{\rm c}=0$ \cite{Neff_Biot07}.}, and the balance of angular momentum equation has to be modified accordingly. All theories in which the stress tensor may not be symmetric can be regarded as polar-continua. The non-symmetry of the stress tensor may also appears also if higher order deformation gradients are included in the free energy, instead of only the first order gradients. Both such theories typically predict a size-effect, meaning that smaller samples of the same material behave relatively stiffer than larger samples. This is an often observed experimental fact, but completely neglected in the classical approach. It implies that some of the additional parameters in the Cosserat model define a length-scale present in the material.

The linear static Cosserat model may be posed in a variational format as a two-field minimization problem for the usual displacement $u$ and the three entries of the infinitesimal microrotation $\mathbf{A}$, which is an element of the Lie-algebra $\so(3)$ of skew symmetric matrices. 
The mathematical analysis of linear micropolar models is fairly well established with a wealth of analytical solutions for boundary value problems, existence and uniqueness theorems and continuous dependence results. It is usually based on a uniform positivity assumption on the free energy which sets it apart from linear elasticity in that Korn‘s inequality is  not needed.

As often the case, notation is a nightmare. Unfortunately, in earlier works of Eringen and others, the interpretation of elastic constants have been misleading with the consequence of giving erroneous parameter ranges for positive definiteness. This has been corrected by S.C. Cowin \cite{Cowin70} and the more recent book by Eringen \cite{Eringen99}.  

Since any skew-symmetric matrix $\mathbf{A}\in \so(3)$ can be identified with its axial vector $\vartheta={\rm axl}(\mathbf{A})$, practically all previous developments for the linear Cosserat model have automatically opted for a presentation of the model in terms of the displacement $u\in \mathbb{R}^3$ and the microrotation vector $\vartheta\in \mathbb{R}^3$. The curvature expression is then simply a quadratic form of the second order tensor ${\rm D}\vartheta \in \mathbb{R}^{3\times 3}$. The advantage of a concise formulation is, however, bought at the expense of transparency of the modelling. Moreover, the advantage of using a microrotation vector is immediately lost when considering a   geometrically  nonlinear Cosserat model in which there appears an orthogonal matrix (the tri\`{e}dre mobil) $\overline{ \mathbf{R}}\in {\rm SO}(3)$. Here, the extraction of a microrotation vector is algebraically  difficult and completely unnatural. Therefore, we advocate a presentation of the linear Cosserat model fully in terms of skew-symmetric matrices $\mathbf{A}\in \so(3)$. The remaining question is how to concisely  express the curvature energy, now naively a quadratic term in the third order tensor ${\rm D}\mathbf{A}\in \mathbb{R}^{3\times 3\times 3}$. However, using ${\rm D}\mathbf{A}$ can be avoided by taking recourse to the so called Nye's formula \cite{Nye53}, expressing  ${\rm D}\vartheta={\rm D}{\rm axl}(\mathbf{A})$ in terms of the matrix ${\rm Curl }\,\mathbf{A}$, i.e.,
\begin{align}\label{curlaxl0}
-\Curl\, \mathbf{A}=(\mathrm{D}\, \axl  \,\mathbf{A})^T-\tr[(\mathrm{D}\, \axl  \,\mathbf{A})^T]{\cdot} \id,\qquad \qquad 
\mathrm{D}\, \axl  \,\mathbf{A}  = -(\Curl\, \mathbf{A})^T+\frac{1}{2}\tr[(\Curl\, \mathbf{A})^T]{\cdot}\id,
\end{align}
and using
\begin{align}
({\rm D}\mathbf{A})_{ijl}=A_{ij,l}=-\epsilon_{ijk} \, {\rm axl}(\mathbf{A})_{k,l}=-\epsilon_{ijk}({\rm D}{\rm axl}(\mathbf{A}))_{kl}.
\end{align}
We note that ${\rm Curl}\,\mathbf{A}$ is a second order tensor which effectively controls all partial derivatives of $\mathbf{A}$, see \cite{Neff_curl06}. For reasons connected to plasticity theory, we will call ${\rm Curl}\, \mathbf{A}$ the dislocation density tensor. Thus, our new Cosserat formulation can be fully expressed in the triplet $(u, \mathbf{A}, {\rm Curl}\, \mathbf{A})$ and the main aim of the current contribution is a complete comparison of the microrotation vector approach in $(u, \vartheta, {\rm D}\, \vartheta)$ versus the new representation. We can already note that in terms of $(u, \mathbf{A}, {\rm Curl}\, \mathbf{A})$ there is a straightforward  extension to the geometrically nonlinear Cosserat model as well as a transparent way to relate to the family of micromorphic models. Especially, the newly developed relaxed micromorphic model practically uses the same curvature expression as our new formulation. The linear Cosserat model can then be obtained as a singular  limit of the relaxed micromorphic model.

Cosserat media may serve as  a model for the prediction of size-effects in foam like structures (like bones) or cellular materials.
This approach has been championed by Lakes \cite{Lakes1,Lakes81,Lakes85b,Lakes95b,lakes2016physical,lakes2018stability,lakes2021softening}. He determines Cosserat parameters by careful size experiments. His values seem to be the only consistent choice of Cosserat parameters ever given for the linear isotropic model. Interestingly, his values have been rejected by a prominent proponent of the micropolar model because Lakes values make the Cosserat free energy only positive semi-definite instead of some supposed positive definiteness in the curvature term. A careful mathematical inspection of analytical solution reveals, however, that Lakes parameter range is a must in order to avoid certain unphysical stiffening behaviour of the Cosserat model for very small samples \cite{neff2010stable,lakes2016physical}.

After a short section introducing the main notations of this paper, in Section \ref{sec:intro} we present the linear  Cosserat model for isotropic elastic materials  in the Eringen's microrotation vector notation. In Section \ref{subsC} we give an alternative formulation of the linear  Cosserat model for isotropic elastic materials in terms of a new strain measure, the dislocation tensor. In Section \ref{idsec} we provide the identification between all constitutive parameters involved in the classical formulation given by Eringen and the constitutive parameters involved in our formulation. In Section \ref{Const} we express some well known constitutive requirements, e.g., positive definiteness of the internal energy density, real plane waves propagation, Legendre-Hadamard ellipticity (strong ellipticity) condition, in terms of the constitutive coefficients considered in our new formulation. In Section \ref{fnot} we compare the constitutive parameters with those used in other formulations, i.e., Nowacki's formulation and Eringen's initial formulation.  In Section \ref{rel} we show that our formulation is a particular case of a formulation given in a more general model, i.e., the relaxed micromorphic model. In Section \ref{Cos} we exhibit that our linear formulation follows directly  from a nonlinear model for isotropic elastic Cosserat solids. In  Section \ref{conclusion} we give a concluding table which facilitates a translation of the experimental results interpreted in terms of the coefficients from other formulation to the constitutive parameters used in our description.

\section{Notation}
We consider that the mechanical behaviour of a body accupying  the unbounded regular region of three dimensional Euclidean space  is modelled with the help of the Cosserat theory of  linear isotropic elastic materials. We denote  by $n$ the outward unit normal on $\partial\Omega$. The body is referred to a fixed system of rectangular Cartesian axes  $Ox_i (i=1,2,3)$ , $\{e_1, e_2, e_3\}$ being the unit vectors of these axes.
Throughout this paper (when we do not specify else) Latin subscripts
take the values $1,2,3$.  Everywhere we adopt the Einstein convention of summation over repeated
indices if not differently specified.

In the following, we recall some useful notations for the present work. For $ a,b\in \mathbb{R}^{3 \times 3} $ we let $\langle{a,b}\rangle_{\mathbb{R}^3}$ denote the scalar product on $ \mathbb{R}^3$ with associated vector norm $\norm{a}^2=\langle {a,a}\rangle$. We denote by $\mathbb{R}^{3\times 3}$   the set of real $3 \times 3$ second order tensors, written with capital letters. Matrices will be denoted by bold symbols, e.g. $\mathbf{X}\in \mathbb{R}^{3\times 3}$, while $X_{ij}$ will denote its component. The standard Euclidean product on $\mathbb{R}^{3 \times 3}$ is given by $\langle{ \mathbf{X},\mathbf{Y}}\rangle_{\mathbb{R}^{3 \times 3}}=\tr( \mathbf{X}\,\mathbf{Y}^T)$, and thus, the Frobenious tensor norm is $\norm{ \mathbf{X}}^2=\langle{ \mathbf{X}, \mathbf{X}}\rangle_{\mathbb{R}^{3 \times 3}}$. In the following we omit the index $\mathbb{R}^3, \mathbb{R}^{3\times 3}$. The identity tensor on $\mathbb{R}^{3\times 3}$ will be denoted by $\id$, so that $\tr( \mathbf{X})=\langle{ \mathbf{X},\id}\rangle$. We let $\Sym$  denote the set of symmetric tensors. We adopt the usual abbreviations of  Lie-algebra theory, i.e., $\mathfrak{so}(3):=\{ \mathbf{A}\,\in \mathbb{R}^{3\times 3}| \mathbf{A}^T=- \mathbf{A}\}$ is the Lie-algebra of skew-symmetric tensors and $\mathfrak{sl}(3):=\{ \mathbf{X}\, \in \mathbb{R}^{3\times 3} |\tr( \mathbf{X})=0 \}$  is the Lie-algebra of traceless tensors. For all $ \mathbf{X} \in\mathbb{R}^{3\times 3}$  we set $\sym\,  \mathbf{X}=\frac{1}{2}( \mathbf{X}^T+ \mathbf{X}) \in\Sym(3),\,\skw  \mathbf{X}=\frac{1}{2}( \mathbf{X}- \mathbf{X}^T)\in \mathfrak{so}(3)$ and the deviatoric (trace-free) part $\dev \, \mathbf{X}= \mathbf{X}-\frac{1}{3}\tr( \mathbf{X})\,\id\in\, \mathfrak{sl}(3)$ and we have the orthogonal Cartan-decomposition of the Lie-algebra $
\mathfrak{gl}(3)=\{\mathfrak{sl}(3)\cap \Sym(3)\}\oplus\mathfrak{so}(3)\oplus\mathbb{R}\cdot\id,$
\begin{align}
\mathbf{X}=\dev  \,\sym\, \mathbf{X}+\skw\, \mathbf{X}+\frac{1}{3}\tr( \mathbf{X})\,\id.\
\end{align}
We use the canonical identification of $\mathbb{R}^3$ with $\so(3)$, and, for
\begin{align}
\mathbf{A}=	\begin{footnotesize}\begin{pmatrix}
0 &-a_3&a_2\\
a_3&0& -a_1\\
-a_2& a_1&0
\end{pmatrix}\end{footnotesize}\in \so(3)
\end{align}
we define the operators $\axl\,:\so(3)\rightarrow\mathbb{R}^3$ and $\anti:\mathbb{R}^3\rightarrow \so(3)$ through
\begin{align}
\axl\, \mathbf{A}:&=\left(
a_1,
a_2,
a_3
\right)^T,\quad \quad  \mathbf{A}.\, v=(\axl\,  \mathbf{A})\times v, \qquad \qquad (\anti(v))_{ij}=-\epsilon_{ijk}\,v_k, \quad \quad \forall \, v\in\mathbb{R}^3,
\notag \\(\axl\,  \mathbf{A})_k&=-\frac{1}{2}\, \epsilon_{ijk} \, \mathbf{A}_{ij}=\frac{1}{2}\,\epsilon_{kij} \, {A}_{ji}\,, \quad  {A}_{ij}=-\epsilon_{ijk}\,(\axl\,  \mathbf{A})_k=:\anti(\axl\,  \mathbf{A})_{ij},
\end{align}
where $\epsilon_{ijk}$ is the totally antisymmetric third order permutation tensor.

For a  regular enough function $f(t,x_1,x_2,x_3)$,  $f_{,t}$ denotes the derivative with respect to the time $t$, while  $ \frac{\partial\, f}{\partial \,x_i}$ and $ f_{,i}$ denotes the $i$-component of the gradient ${\rm D} f$.  For vector fields $u=\left(    u_1, u_2, u_3\right)^T$ with  $u_i\in 
{\rm H}^1(\Omega)\,=\,\{u_i\in {\rm L}^2(\Omega)\, |\, {\rm D} u_i\in {\rm L}^2(\Omega)\}, $  $i=1,2,3$,
we define
$
\mathrm{D} u:=\left(
{\rm D}  u_1\,|\,
{\rm D} u_2\,|\,
{\rm D} u_3
\right)^T.
$

The corresponding Sobolev-space will be also denoted by
$
{\rm H}^1(\Omega)$.  For vector fields $u$ with components in ${\rm H}^{1}(\Omega)$ and tensor fields $P$ with rows in ${\rm H}({\rm curl}\,; \Omega)$, resp. ${\rm H}({\rm div}\,; \Omega)$, i.e.,
\begin{align}
u=\left(
\begin{array}{c}
u_1 \\
u_2 \\
u_3 \\
\end{array}
\right)\, , u_i\in {\rm H}^{1}(\Omega),
\ \quad
P=\left(
\begin{array}{ccc}
P_{11} &P_{12}&P_{13}\\
P_{21} &P_{22}&P_{23}\\
P_{31} &P_{32}&P_{33}
\end{array}
\right)\, \quad P^Te_i\in {\rm H}({\rm curl}\,; \Omega)\,  \quad \ \text{resp.} \quad P^Te_i\in {\rm H}({\rm div}\,; \Omega)
\end{align}
we define
\begin{align}
{\rm curl}\, u&=\left(
\begin{array}{c}
\dd u_{3,2}-  u_{2,3}\\ \dd  u_{1,3}-\dd  u_{3,1}  \\ \dd u_{2,3}- u_{3,1}
\end{array}
\right)=\left(\epsilon_{ijk}u_{k,j}\right)_{i=1,2,3}, \qquad 
{\rm D}u=
\dd\begin{pmatrix}
\dd  u_{1,1}&\dd u_{1,2}&\dd  u_{1,3}\vspace{1.2mm}\\
\dd u_{2,1}&\dd  u_{2,2}&\dd  u_{2,3}\vspace{1.2mm}\\
\dd  u_{3,1}&  u_{3,2}&\dd  u_{3,3}
\end{pmatrix}=\left(u_{i,j}\right)_{i,j=1,2,3}\, ,\\
\ \ \ \ {\rm Curl}\,\mathbf{P}&=\begin{pmatrix}
[{\rm curl}\, \begin{pmatrix}P_{11} &P_{12}&P_{13}\end{pmatrix}^T]^T\\
[{\rm curl}\, \begin{pmatrix}P_{21} &P_{22}&P_{23}\end{pmatrix}^T]^T\\
[{\rm curl}\, \begin{pmatrix}P_{31} &P_{32}&P_{33}\end{pmatrix}^T]^T\\
\end{pmatrix}=\begin{pmatrix}
\dd P_{13,2}-P_{12,3}& \dd P_{11,3}- P_{13,1}  & \dd P_{12,3}- P_{13,1}\\
\dd P_{23,2}-P_{22,3}& \dd P_{21,3}-\dd  P_{23,1}  & \dd  P_{22,3}-P_{23,1}\\
\dd P_{33,2}- P_{32,3}& \dd P_{31,3}-\dd P_{33,1}  & \dd  P_{32,3}- P_{33,1}
\end{pmatrix}=\left(\epsilon_{jmn}P_{in,m}\right)_{i,j=1,2,3},\,\notag\\
 {\rm Div}\,\mathbf{P}&=\begin{pmatrix}
 {\rm div}\, \begin{pmatrix}P_{11} &P_{12}&P_{13}\end{pmatrix}^T\\
 {\rm div}\,  \begin{pmatrix}P_{21} &P_{22}&P_{23}\end{pmatrix}^T\\
 {\rm div}\,  \begin{pmatrix}P_{31} &P_{32}&P_{33}\end{pmatrix}^T\\
 \end{pmatrix}=
 \begin{pmatrix}
 	\dd P_{11,1} + \dd P_{12,2} +\dd  P_{13,3} \vspace{1.2mm}\\
 	\dd P_{21,1} + \dd P_{22,2}+\dd  P_{23,3} \vspace{1.2mm}\\
 	\dd P_{31,1} + \dd P_{32,2}+\dd P_{33,3}
 \end{pmatrix}=\left(P_{ij,j}\right)_{i,j=1,2,3}.\notag
\end{align}

\section{Classical linear Cosserat model in Eringen  notation}\setcounter{equation}{0}
\label{sec:intro}
In the isotropic linear Cosserat model the kinematics is described through a vector field,  the displacement $u:[0,T]\times\Omega \subset \mathbb{R}^3 \to \mathbb{R}^3$, and a skew-symmetric tensor, the micro-rotation tensor $\mathbf{A}:[0,T]\times\Omega \subset \mathbb{R}^3 \to \mathfrak{so}(3)$. Being skew-symmetric, the micro-rotation tensor of $\mathbf{A}$ is fully determined by its axial vector, i.e.,
\begin{align}
\vartheta:= \text{axl}\,\mathbf{A}  \, ,
\qquad\qquad 
\textbf{Anti} \, \vartheta = \mathbf{A} \, ,
\end{align}
which is called  the {\it micro-rotation vector}. 

The linear strain tensors considered in the Eringen formulation \cite[Page 104]{Eringen99} are
\begin{align}
{\rm e}^*_{ji} = u_{i,j}+  \epsilon_{ijm}\,\vartheta_m, \qquad\qquad  \mathfrak{K}_{ji} =\vartheta_{j,i}.
\end{align}
 In a compact matrix form these strain tensors are
\begin{align}
{\mathbf{e}^*} =({\rm D}u)^T+\mathbf{ Anti}\,\vartheta =({\rm D}u)^T+\mathbf{A},\qquad\qquad  \boldsymbol{\mathfrak{K}} ={\rm D} \vartheta={\rm D} (\axl \mathbf{A}).
\end{align}

Following Eringen \cite[Page 111]{Eringen99}, the internal density energy for isotropic materials is 
\begin{align}
W =  \frac{1}{2}\Big[& (\mu^*+\varkappa)\, {\rm e}^* _{ji}{\rm e}^* _{ji}+ \mu^*\, {\rm e}^*_{ji} {\rm e}^*_{ij} +\lambda\, {\rm e}^*_{ii}  {\rm e}^*_{jj} +\gamma \, \mathfrak{K}_{ji}\mathfrak{K}_{ji} +\beta\, \mathfrak{K}_{ji} \mathfrak{K}_{ij} + \alpha\, \mathfrak{K}_{ii} \mathfrak{K}_{jj} \Big].
\end{align}
In a compact form, without involving the notations in indices, the expression of the internal density energy is
\begin{align}
W= \frac{1}{2}\Big[&(\mu^*+\varkappa) \bigl\langle {\mathbf{e}^*} ,{\mathbf{e}^*} \bigr\rangle+ \mu^*\bigl\langle {\mathbf{e}^*} ,{\mathbf{e}^*}^T\bigr\rangle+\lambda \left(\tr( {\mathbf{e}^*}) \right)^2+\gamma \bigl\langle \boldsymbol{\mathfrak{K}} ,\boldsymbol{\mathfrak{K}} \bigr\rangle+\beta  \bigl\langle \boldsymbol{\mathfrak{K}} ,\boldsymbol{\mathfrak{K}}^T\bigr\rangle+ \alpha  \left(\tr( \boldsymbol{\mathfrak{K}}) \right)^2\Big].
\end{align}
Let us remark that 
\begin{align}
	W= \frac{1}{2}\Big[&(\mu^*+\varkappa) \bigl\langle (\sym \,{\mathbf{e}^*} +\skw \, {\mathbf{e}^*} ) ,(\sym \,{\mathbf{e}^*} +\skw \, {\mathbf{e}^*} ) \bigr\rangle\notag
	\\&+ \mu^*\bigl\langle (\sym \,{\mathbf{e}^*} +\skw \, {\mathbf{e}^*} ) ,(\sym \,{\mathbf{e}^*} -\skw \, {\mathbf{e}^*} \bigr\rangle+\lambda \left(\tr (\sym \,{\mathbf{e}^*} ) \right)^2\notag\\&+\gamma  \bigl\langle (\sym \, \boldsymbol{\mathfrak{K}}+\skw \, \boldsymbol{\mathfrak{K}}) ,(\sym \, \boldsymbol{\mathfrak{K}}+\skw \, \boldsymbol{\mathfrak{K}}) \bigr\rangle\notag\\&+\beta  \bigl\langle (\sym \, \boldsymbol{\mathfrak{K}}+\skw \, \boldsymbol{\mathfrak{K}}) ,(\sym \, \boldsymbol{\mathfrak{K}}-\skw \, \boldsymbol{\mathfrak{K}})\bigr\rangle+ \alpha  \left(\tr (\sym \, \boldsymbol{\mathfrak{K}}) \right)^2\Big]\\=\frac{1}{2}\Big[&(\mu^*+\varkappa) \| \sym \,{\mathbf{e}^*} \|^2+(\mu^*+\varkappa) \|\skw \, {\mathbf{e}^*}  \|^2\notag
+ \mu^*\|\sym \,{\mathbf{e}^*} \|^2-\mu^*\|\skw \, {\mathbf{e}^*} \|^2+\lambda \left(\tr (\sym \,{\mathbf{e}^*} ) \right)^2\notag\\&+\gamma \, \|\sym \, \boldsymbol{\mathfrak{K}}\|^2+\gamma\,\|\skw \, \boldsymbol{\mathfrak{K}}\|^2+\beta\,  \|\sym \, \boldsymbol{\mathfrak{K}}\|^2-\beta\, \|\skw \, \boldsymbol{\mathfrak{K}}\|^2+ \alpha\,  \left(\tr (\sym \, \boldsymbol{\mathfrak{K}}) \right)^2\Big]\notag
\\=\frac{1}{2}\Big[&(2\,\mu^*+\varkappa) \| \sym \,{\mathbf{e}^*} \|^2+\varkappa\,\|\skw \, {\mathbf{e}^*}  \|^2+\lambda \left(\tr (\sym \,{\mathbf{e}^*} ) \right)^2\notag\\&+(\gamma+\beta )  \|\sym \, \boldsymbol{\mathfrak{K}}\|^2+(\gamma-\beta ) \|\skw \, \boldsymbol{\mathfrak{K}}\|^2+ \alpha  \left(\tr (\sym \, \boldsymbol{\mathfrak{K}}) \right)^2\Big].\notag
\\=\frac{1}{2}\Big[&(2\,\mu^*+\varkappa) \| \sym \,\text{D}u\|^2+\varkappa\,\|\skw\, {\rm D} u-\mathbf{A}\|^2+\lambda \left(\tr (\text{D}u ) \right)^2\notag\\&+(\gamma+\beta )  \|\sym \, \text{D}\vartheta\|^2+(\gamma-\beta ) \|\skw \, \text{D}\vartheta\|^2+ \alpha  \left(\tr (\text{D}\vartheta) \right)^2\Big].\notag
\end{align}
Since
\begin{align}
2\,  \axl (\skw\, {\rm D} u)&={\rm curl}\, u, \qquad 2\,   (\skw\, {\rm D} u)=\mathbf{Anti}( {\rm curl}\, u),
\end{align}
and 
\begin{align}
\|\skw\, {\rm D} u-\mathbf{A}\|^2&=2\|{\rm axl}(\skw\, {\rm D} u-\mathbf{Anti} \vartheta)\|^2=2\,\|\frac{1}{2}{\rm curl}\,u-\vartheta\|^2=\frac{1}{2}\,\|{\rm curl}\,u-2\,\vartheta\|^2,\\
\|\skw\, {\rm D} \vartheta\|^2&=2\|{\rm axl}(\skw\, {\rm D} \vartheta\|^2=2\,\|\frac{1}{2}{\rm curl}\,\vartheta\|^2=\frac{1}{2}\,\|{\rm curl}\,\vartheta\|^2,\notag\\
\|\boldsymbol{\mathfrak{K}}\|^2&=\|{\rm D} \vartheta\|^2=\|{\rm D}{\rm axl}(\mathbf{A}) \|^2=\frac{1}{2}\,\|{\rm D}\mathbf{A}\|^2\notag
\end{align}
we obtain the following equivalent alternative form of the internal energy density 
\begin{align}\label{energyforms}
W= \Big[&(\mu^*+\frac{\varkappa}{2}) \| \sym \,\text{D}u\|^2+\frac{\varkappa}{4}\,\|{\rm curl}\,u-2\,\vartheta\|^2+\frac{\lambda}{2} \left(\tr (\text{D}u ) \right)^2\notag\\&+\frac{(\gamma+\beta )}{2}  \|\sym \, \text{D}\vartheta\|^2+\frac{(\gamma-\beta )}{4} \,\|{\rm curl}\,\vartheta\|^2+ \frac{\alpha}{2}   \left(\tr (\text{D}\vartheta) \right)^2\Big]\\
=\Big[&(\mu^*+\frac{\varkappa}{2}) \| \sym \,\text{D}u\|^2+\frac{\varkappa}{4}\,\|{\rm curl}\,u-2\,\vartheta\|^2+\frac{\lambda}{2} \left(\tr (\text{D}u ) \right)^2\notag\\&+\frac{(\gamma+\beta )}{2}  \|\dev\,\sym \, \text{D}\vartheta\|^2+\frac{(\gamma-\beta )}{4} \,\|{\rm curl}\,\vartheta\|^2+ \frac{ {3\,\alpha+}\gamma+\beta }{6}   \left(\tr (\text{D}\vartheta) \right)^2\Big]\notag\\
=\Big[&(\mu^*+\frac{\varkappa}{2}) \| \sym \,\text{D}u\|^2+\frac{\varkappa}{4}\,\|{\rm curl}\,u-2\,\vartheta\|^2+\frac{\lambda}{2} \left( {{\rm div}\,u } \right)^2\notag\\&+\frac{(\gamma+\beta )}{2}  \|\dev\,\sym \, \text{D}\vartheta\|^2+\frac{(\gamma-\beta )}{4} \,\|{\rm curl}\,\vartheta\|^2+ \frac{ {3\,\alpha+}\gamma+\beta}{6}   \left( {{\rm div}\,\vartheta} \right)^2\Big].\notag
\end{align}

In indices notation \cite[Page 111]{Eringen99} we have the following expression of 
the non-symmetric force stress and the couple stress tensor definition  
\begin{align}\label{eq:stress_classic}
{\sigma} _{ji} &=\frac{\partial W}{\partial {\rm e}^*_{ji} }=
(\mu^*+\varkappa)  \,{\rm e}^* _{ji}+ \mu^*{\rm e}^* _{ij}+\lambda \,   {\rm e}^* _{kk}\delta_{ji} =
(\mu^*+\varkappa)  (u_{i,j}+  \epsilon_{ijm}\vartheta_m)+ \mu^*( u_{j,i}+  \epsilon_{jim}\vartheta_m)+\lambda \,    u_{k,k}\delta_{ji},\notag
\\
m _{ji}^*&=\frac{\partial W}{\partial \mathfrak{K}_{ji} }=
\beta  \, \mathfrak{K}_{ij} +\gamma\,  \mathfrak{K}_{ji}+ \alpha \,   \mathfrak{K}_{kk}\delta_{ji} =
\beta \,  \vartheta_{j,i} +\gamma \, \vartheta_{i,j}+ \alpha   \,  \vartheta_{k,k}\delta_{ji} 
\, .
\end{align}
In a  compact matrix description, 
the force stress and the couple stress tensor definition  in the \textit{classical form} are given by\footnote{These expressions agree with  the form considered by Eremeyev  et al.   \cite[Eq.~(4.68)]{eremeyev2012foundations}, where the internal energy is written as
\begin{align}
W=&\frac{\lambda}{2}\left[\tr(\mathbf{\varepsilon})\right]^2+\frac{\mu^*+\varkappa}{2} \tr(\mathbf{\varepsilon}\, \mathbf{\varepsilon}^T)+\frac{\mu^*}{2} \tr(\mathbf{\varepsilon}\, \mathbf{\varepsilon})+\frac{\beta_1}{2}\left[\tr(\boldsymbol{\mathfrak{K}})\right]^2+\frac{\beta_2}{2} \tr(\boldsymbol{\mathfrak{K}}\, \boldsymbol{\mathfrak{K}}^T)+\frac{\beta_3}{2} \tr(\boldsymbol{\mathfrak{K}}\, \boldsymbol{\mathfrak{K}})
.\end{align}}
\begin{align}\label{eq:stress_classic}
\boldsymbol{\sigma} &={\rm D}_{{\mathbf{e}^*} }W=
(\mu^*+\varkappa) \, {\mathbf{e}^*} + \mu^*{\mathbf{e}^*}^T+\lambda \tr (\mathbf{e}^*) \,\id\notag\\&=
(\mu^*+\varkappa) \, (\sym \,{\mathbf{e}^*} +\skw \, {\mathbf{e}^*} ) +\mu^*(\sym\, {\mathbf{e}^*} -\skw \, {\mathbf{e}^*} )+\lambda \tr( {\mathbf{e}^*}) \,\id\notag\\&=
(2\, \mu^*+\varkappa)\,\sym \, {\mathbf{e}^*} + \varkappa\,  \skw\, {\mathbf{e}^*} +\lambda \tr ({\mathbf{e}^*}) \,\id\\&=
(2\, \mu^*+\varkappa)\,\sym \,  {\rm D}u+ \varkappa  (\skw (  {\rm D}u)^T+\mathbf{Anti}\,\vartheta)+\lambda \tr   ({\rm D}u)\,\id\,
\notag\\&=
(2\, \mu^*+\varkappa)\,\sym \,  {\rm D}u- \varkappa  (\skw (  {\rm D}u)-\mathbf{Anti}\,\vartheta)+\lambda \tr   ({\rm D}u)\,\id\, 
\notag\\&=
(2\, \mu^*+\varkappa)\,\sym \,  {\rm D}u- \frac{\varkappa}{2} \, \mathbf{Anti}({\rm curl}\,u-2\,\vartheta)+\lambda \tr   ({\rm D}u)\,\id\, ,\notag
\\*
\boldsymbol{m}^*  &={\rm D}_{\boldsymbol{\mathfrak{K}} }W=\gamma \, \boldsymbol{\mathfrak{K}}+
\beta \,  \boldsymbol{\mathfrak{K}}^T + \alpha\,  \tr (\boldsymbol{\mathfrak{K}}) \, \id\notag=
\beta  \,   {\rm D}\vartheta +\gamma\,  (  {\rm D}\vartheta )^T+ \alpha\,  \tr   ({\rm D}\vartheta) \, \id\notag
\, .
\notag
\end{align}

By considering  the action functional 
\begin{align}
\mathcal{A}=&\int_{0}^{T}\int_\Omega\bigg(\frac{1}{2}\,\rho\,\|\dot{u}\|^2+\,\rho\,j\,\|\dot{\vartheta}\|^2
-W)\, dv\,dt,
\end{align}
where $j$ is a micro-inertia coefficient,  we obtain in  the absence of body loads, the motion equations in the \textit{classical form} 
\begin{align}
 \varrho \, {u}_{i,tt}&=\sigma_{ji,j} , \qquad \qquad \qquad \ \, i=1,2,3,\\
  \varrho \,j\, {\vartheta}_{i,tt}&=m_{ji,j}^* +  \epsilon_{ijk}\,\sigma_{jk} , \qquad i=1,2,3,\notag
\end{align}
equivalently, in  matrix (\textbf{Div-Div}) form,
\begin{align}
\varrho \, {u}_{,tt}&=\text{Div}\,\boldsymbol{\sigma}^T,\\
 \varrho \,j\, {\vartheta}_{,tt}&=\text{Div}\, \boldsymbol{m}^{*T} -2\,\axl\,(\skw\, \boldsymbol{\sigma}).\notag
\end{align}

In the absence of  external  body forces and   external  body moment,  the PDE-system of the model \cite[Pages 118-119]{Eringen99} is 
\begin{align}\label{PDE}
\rho\,u_{i,tt}&=(\mu^*+\varkappa )\,\dd  u_{i,jj} +(\lambda+\mu^*)\,\dd  u_{j,ij}+\varkappa  \,  \epsilon _{ijk}\,\vartheta _{k,j} ,  \qquad \ i=1,2,3
\vspace{1.2mm}\\
\rho\, j\,\vartheta_{i,tt}&=(\alpha +\beta )\,\dd  \vartheta _{i,ij}+\gamma\, \dd   \vartheta_{i,jj}+\,\varkappa \,  \epsilon _{ijk} u_{k,j} -2\, \varkappa\, \vartheta _i, \qquad i=1,2,3. \notag
\end{align}

 In a compact form,  the above PDE-system reads 
\begin{align}\label{PDE2}
\rho\,\frac{\partial^2 \,u}{\partial\, t^2}&=(\mu^*+\varkappa )\,\dd  \Delta u +(\mu^*+\lambda)\,\dd  {\rm D} ({\rm div}\, u)+{\varkappa}  \, {\rm curl} \, \vartheta, \qquad\  i=1,2,3,
\vspace{1.2mm}\\
\rho\, j\,\frac{\partial^2 \,\vartheta}{\partial\, t^2}&=\gamma \,\dd  \Delta \vartheta+(\alpha +\beta )\, \,\dd  {\rm D} ({\rm div}\, \vartheta)+\,{\varkappa}\,\left( \,  {\rm curl}\, u-2\, \vartheta\right),  \qquad i=1,2,3.  \notag
\end{align}
Using the well-known identity
\begin{align}
0\ =\ - {\rm curl} \left(  {\rm curl}\,u \right) +{\rm D}({\rm div} \,u) \,-\, \Delta u
\end{align}
the PDE-system can be rewritten as
\begin{align}\label{PDE22}
\rho\,\frac{\partial^2 \,u}{\partial\, t^2}&=\mu^*\dd  \Delta u +(\mu^*+\lambda+\varkappa)\dd  {\rm D} ({\rm div}\, u)-\varkappa\,{\rm curl}\, \left(   {\rm curl} \,u  - \vartheta\right),
\vspace{1.2mm}\\
\rho\, j\,\frac{\partial^2 \,\vartheta_i}{\partial\, t^2}&=\gamma \,\dd  \Delta \vartheta+(\alpha +\beta )\, \,\dd  {\rm D} ({\rm div}\, \vartheta)+\,{\varkappa}\left( {\rm curl}\, u-2\, \vartheta\right).  \notag\\
\notag
\end{align}
\section{Cosserat theory of isotropic elastic solids in dislocation format}\label{subsC}\setcounter{equation}{0}

Now we propose another form of the internal energy density and we show that after an identification of the parameters it coincides with the internal energy density proposed by Eringen. However, the new form of the internal energy density has the advantage to have the same structure as in the case of a more generalised theory, i.e., the relaxed micromorphic model \cite{NeffGhibaMicroModel,MadeoNeffGhibaW,MadeoNeffGhibaWZAMM,madeo2016reflection,NeffGhibaMadeoLazar}, see Subsection \ref{rel}  and the nonlinear Cosserat model \cite{neff2015existence}, see Subsection  \ref{Cos}.

We keep the physical meaning of the unknowns $u$ and $\vartheta={\rm axl}(\mathbf{A})$ from the Cosserat theory,  but we propose a new form of the Cosserat internal energy density, namely
\begin{align}\label{XXXX}
W&=\mu_{\rm e} \|\sym \,(\mathrm{D}u -\mathbf{A})\|^2+\mu_{\rm c} \|\skw(\mathrm{D}u -\mathbf{A})\|^2+ \frac{\lambda_{\rm e} }{2}\, [\tr(\mathrm{D}u -\mathbf{A})]^2\notag\\&
\quad \quad  +\frac{\mu_{\rm e} L_{\rm c}^2}{2}\left[{a_1}\|\dev  \,\sym \,\Curl\, \mathbf{A}\|^2 +{a_2}\| \skw \,\Curl\, \mathbf{A}\|+ \frac{a_3}{3}\, \tr(\Curl\, \mathbf{A})^2\right],
\end{align}
where $(\mu_{\rm e} ,\lambda_{\rm e} ) $,   $\mu_{\rm c}, L_{\rm c} $  and $( a_1, a_2, a_3)$  are the elastic moduli  representing the parameters related to the meso-scale, the  Cosserat couple modulus $\mu_{\rm c}$, the characteristic length $L_{\rm c} $, and the three
general isotropic curvature parameters (nondimensional weights), respectively.

The new energy is expressed in terms of the strain and curvature  tensors
\begin{align}
\mathbf{e}=  {\rm D}u-\mathbf{Anti}\,\vartheta =  {\rm D}u-\mathbf{A}\in\mathbb{R}^{3\times 3},\qquad\qquad  \boldsymbol{\alpha}:=-\Curl \, \mathbf{A}\in\mathbb{R}^{3\times 3}.
\end{align}

The stress and moment definition in the \textit{dislocation form} are
\begin{align}
\boldsymbol{\sigma}&={\rm D}_{\mathbf{e}}W
=
 2\mu_{\rm e} \,\text{sym} \, \text{D}u
+ 2\mu_{\rm c}\,\text{skew} \, \text{D}u - \mathbf{A}+\lambda_{\rm e}  \, \text{tr} \left(\text{D}u \right) \boldsymbol{\mathbbm{1}}
\, \notag\\
&=
2\mu_{\rm e} \,\sym \,  {\rm D}u+ 2\mu_{\rm c} (\skw (  {\rm D}u)-\mathbf{Anti}\,\vartheta)+\lambda_{\rm e}\, \tr   ({\rm D}u)\,\id\,
\notag\\&=
2\mu_{\rm e} \,\sym \,  {\rm D}u+\mu_{\rm c} \, \mathbf{Anti}({\rm curl}\,u-2\,\vartheta)+\lambda_{\rm e}\, \tr   ({\rm D}u)\,\id\, ,
\label{eq:stress_dislo}
\\*
\boldsymbol{m} &={\rm D}_{\boldsymbol{\alpha}}W=\mu_{\rm e} 
L_{\rm c}^2 \, 
\left(
a_1 \, \text{dev} \, \text{sym} \, \text{Curl} \, \mathbf{A}
+ a_2 \, \text{skew} \, \text{Curl} \, \mathbf{A}
+ \frac{a_3}{3} \, \text{tr} \left(\text{Curl} \, \mathbf{A} \right)\boldsymbol{\mathbbm{1}} \, 
\right) \, .
\notag
\end{align}

Considering the action functional 
\begin{align}
\mathcal{A}=&
\int_{0}^{T}\int_\Omega\bigg(\frac{1}{2}\,\rho\,\|u_{,t}\|^2+\frac{1}{2}\,\rho\,\eta\,\tau_{\rm c}^2\,\,\,\| \,\mathbf{A}_{,t}\|^2
-W\bigg)\, dv\,dt
\end{align}
with  $\eta\,\tau_{\rm c}^2$ an inertia coefficient, $\eta>0$ a nondimensional weight parameter and $\tau_{\rm c}$ the internal characteristic time \cite[page 163]{Eringen99}, we obtain  the Euler-Lagrange equation, i.e., the following system in \textbf{Div-Curl-form} (the dislocation formulation)

\begin{align}
\rho\,u_{,tt}&=\text{Div}\, \boldsymbol{\sigma} \, ,
\label{eq:equi_Cos}
\\
\rho\eta\, \tau_{\rm c}^2\,\mathbf{A}_{,tt}&=
- \text{skew} \, \text{Curl}\,
\boldsymbol{m} +\text{skew} \,\boldsymbol{\sigma}\,,
\notag
\end{align}
which, expressed in terms of  $u$ and $\mathbf{A}$ gives
\begin{align}
\rho\,u_{,tt}&=\text{Div}\left[ 2\mu_{\rm e}\,\text{sym} \, \text{D}u
+ 2\mu_{\text{c}}\,\text{skew} \left(\text{D}u - \mathbf{A}\right)+
\lambda_{\rm e} \, \text{tr} \left(\text{D}u \right) \boldsymbol{\mathbbm{1}}
\right] \, ,
\label{eq:equi_Cos}
\\
\rho\eta\, \tau_{\rm c}^2\,\mathbf{A}_{,tt}&=
-L_{\rm c}^2 \,  \, \mu_{\rm e}\,\text{skew} \, \text{Curl}\,
\left[
a_1 \, \text{dev} \, \text{sym} \, \text{Curl} \, \mathbf{A}
+ a_2 \, \text{skew} \, \text{Curl} \, \mathbf{A} 
+ \frac{a_3}{3} \, \text{tr} \left(\text{Curl} \, \mathbf{A} \right)\boldsymbol{\mathbbm{1}} \, 
\right] +2\mu_{\text{c}}\,\text{skew} \left(\text{D}u - \mathbf{A}\right)\,.
\notag
\end{align}

We recall 
the  {\bf Curl-$\boldsymbol{\mathrm{D}\,\axl}\,$ identitie}s,  (see \cite{Neff_curl06}, {\bf Nye's formula} \cite{Nye53})
\begin{align}\label{curlaxl}
\boldsymbol{\alpha}&=-\Curl\, \mathbf{A}=(\mathrm{D}\, \axl  \,\mathbf{A})^T-\tr[(\mathrm{D}\, \axl  \,\mathbf{A})^T]{\cdot} \id=\boldsymbol{\mathfrak{K}}^T-\tr(\boldsymbol{\mathfrak{K}}^T)\, \id,\\
\boldsymbol{\mathfrak{K}} &=\mathrm{D}\, \axl  \,\mathbf{A}  = -(\Curl\, \mathbf{A})^T+\frac{1}{2}\tr[(\Curl\, \mathbf{A})^T]{\cdot}\id=\boldsymbol{\alpha}^T-\frac{1}{2}\tr(\boldsymbol{\alpha}^T)\,\id,\notag
\end{align}
as well as its implications
\begin{align}\label{curlaxli}
\sym\,\boldsymbol{\alpha}&=\sym\,\boldsymbol{\mathfrak{K}}-\tr(\boldsymbol{\mathfrak{K}})\, \id,\qquad \qquad 
\dev\,\sym\,\boldsymbol{\alpha}=\dev\,\sym\,\boldsymbol{\mathfrak{K}},\\
\skw\,\boldsymbol{\alpha}&=-\skw \, \boldsymbol{\mathfrak{K}},\qquad\qquad \qquad \qquad \quad\ \,
\tr(\boldsymbol{\alpha})=-2\,\tr(\boldsymbol{\mathfrak{K}}).\notag
\end{align}

Due to this converting formulae,  the action functional admits the equivalent form
\begin{align}
\mathcal{A}
=&\int_{0}^{T}\int_\Omega\bigg(\frac{1}{2}\,\rho\,\|u_{,t}\|^2+\,\rho\,\eta\,\tau_{\rm c}^2\,\,\,\|(\axl\, \,\mathbf{A})_{,t}\|^2
-W\bigg)\, dv\,dt\notag,
\end{align}
where the internal energy density is expressed in  the alternative form
\begin{align}
W=& \,\mu_{\rm e} \|\sym \,(\mathrm{D}u -\mathbf{A})\|^2+\mu_{\rm c} \|\skw(\mathrm{D}u -\mathbf{A})\|^2+ \frac{\lambda_{\rm e} }{2}\, [\tr(\mathrm{D}u -\mathbf{A})]^2\notag\\& +\frac{\mu_{\rm e} L_{\rm c}^2}{2}\,\left[a_1\,\lVert\dev   \,\sym\,\mathrm{D}\, \axl  \,\mathbf{A}  \rVert ^{2}+a_2\,\lVert \skw\,\mathrm{D}\, \axl  \,\mathbf{A}  \rVert ^{2}+\frac{4\,a_3}{3}\left[\mathrm{tr} \left(\mathrm{D}\, \axl  \,\mathbf{A}  \right)\right]^{2}\right].
\end{align}

We also notice that
\begin{align}
\|\boldsymbol{\mathfrak{K}}\|^2=\|\dev \,\sym \, \boldsymbol{\mathfrak{K}}\|^2+\|\skw \, \boldsymbol{\mathfrak{K}}\|^2+\frac{1}{3}[\tr \boldsymbol{\mathfrak{K}}]^2= \,\|\dev \,\text{sym} \, \boldsymbol{\alpha} \|^2+\,\|\text{skew} \, \boldsymbol{\alpha} \|^2+\frac{1}{12}\,
[{\rm tr}(\boldsymbol{\alpha} )]^2.
\end{align}

Since ${\rm H}^1(\Omega)\cap \mathfrak{so}(3)$ coincides with  ${\rm H}({\rm Curl};\Omega)\cap \mathfrak{so}(3)$, see \cite{Neff_curl06},  taking variations over all $\mathbf{A}\in {\rm H}^1(\Omega)\cap \mathfrak{so}(3)$ is equivalent to considering variations over all $\mathbf{A}\in {\rm H}({\rm Curl};\Omega)\cap \mathfrak{so}(3)$, i.e. the obtained Euler-Lagrange equations obtained using variations in $\mathbf{A}\in {\rm H}^1(\Omega)\cap \mathfrak{so}(3)$ and  $\mathbf{A}\in {\rm H}({\rm Curl};\Omega)\cap \mathfrak{so}(3)$, respectively, are equivalent. Taking variations ${\rm H}({\rm Curl};\Omega)\cap \mathfrak{so}(3)$, 
the minimum action principle leads to  the following system of partial differential equations
\begin{align}\label{eqisaxl}
\rho\,u_{,tt}&=\Div[\underbrace{2\mu_{\rm e} \, \sym\,\mathrm{D}u +2\,\mu_{\rm c} \,\skw(\mathrm{D}u -\mathbf{A})+ \lambda_{\rm e} \, \tr(\mathrm{D}u ){\cdot} }_{\sigma}\id]\, ,\notag\\
2\,\rho\,\eta\,\tau_{\rm c}^2\,\,\,(\axl\, \,\mathbf{A})_{,tt}&=\frac{\mu_{\rm e} L_{\rm c}^2}{2}\,\Div\bigg[2\,{a_1 }\,\dev\,\sym \,(\mathrm{D}\, \axl  \,\mathbf{A})+2\,{a_2}\,\skw (\mathrm{D}\, \axl  \,\mathbf{A})+ \frac{8\,a_3}{3}\,\tr(\mathrm{D}\, \axl  \,\mathbf{A}){\cdot } \id\bigg]
\\&\ \ \ \ \ \ \ \ \ \ \ \ \ -4\,\mu_{\rm c} \,\axl\,(\skw\,\mathrm{D}u -\mathbf{A})\, ,\notag
\end{align}
 The above systems is in complete agreement to the equations proposed by Eringen in the linear Cosserat theory.  To see this explicitly, let us write this PDE-system  in indices, giving
\begin{align}\label{PDE}
\rho\,\,u_{i,tt}&=(\mu_{\rm e} +\mu_{\rm c})\dd   u_{i,jj}+(\mu_{\rm e} -\mu_{\rm c}+\lambda_{\rm e}  )\dd  u_{j,ij}+2\,\mu_{\rm c} \,  \epsilon _{ijk}\vartheta _{k,j},
\vspace{1.2mm}\\
2\,\rho\, \eta\,\mu_{\rm e} \,\tau_{\rm c}^2\,\vartheta_{i,tt}&=\mu_{\rm e} \frac{L_{\rm c}^2}{2}\,\left[(a_1+a_2)\,  \vartheta _{i,jj} +\frac{1}{3}\,(a_1-3\,a_2+2\,a_3))\vartheta _{j,ij}\right]+2\,\mu_{\rm c} \,  \varepsilon _{ijk}u_{k,j}-4\, \mu_{\rm c}\, \vartheta _i \, .
\notag
\end{align}

Other alternative expressions of the energy, in which the quadratic form in terms of $\mathbf{e}$ has a similar structure as the quadratic form in terms of $\mathbf{\mathfrak{K}}$, are
 \begin{align}\label{enersplit}
W=&\underbrace{\mu_{\rm e} \,\lVert\dev   \, \sym\,\mathbf{e}\rVert ^{2}+\mu_{\rm c}\,\lVert \skw\,\mathbf{e}\rVert ^{2}+\frac{2\mu_{\rm e} +3\,\lambda_{\rm e} }{6}\left[\mathrm{tr} \left(\mathbf{e}\right)\right]^{2}}_{:=\,W_1(\mathbf{e})}\notag\\& +\underbrace{\frac{\mu_{\rm e} L_{\rm c}^2}{2}\,\left[\ {\alpha_1}\,\lVert\dev   \,\sym\,\mathbf{\mathfrak{K}}\rVert ^{2}+{\alpha_2}\,\lVert \skw\,\mathbf{\mathfrak{K}}\rVert ^{2}+\frac{2\,\alpha_1+3\,\alpha_3}{6}\left[\mathrm{tr} \left(\mathbf{\mathfrak{K}}\right)\right]^{2}\right]}_{:=\,W_2(\mathbf{\mathfrak{K}})}\\
=&\, \mu_{\rm e} \|\sym \,(\mathrm{D}u -\mathbf{A})\|^2+\mu_{\rm c} \|\skw(\mathrm{D}u -\mathbf{A})\|^2+ \frac{\lambda_{\rm e} }{2}\, [\tr(\mathrm{D}u -\mathbf{A})]^2\notag\\&
\quad \quad  +\frac{\mu_{\rm e} L_{\rm c}^2}{2}\left[{\alpha_1}\|\dev \sym \,\Curl\, \mathbf{A}\|^2 +{\alpha_2}\| \skw \,\Curl\, \mathbf{A}\|+ \frac{2\,\alpha_1+3\,\alpha_3}{24}\, [\tr(\Curl\, \mathbf{A})]^2\right],\notag
\end{align}
where  we have used the  new   weight parameters $\alpha_1, \alpha_2, \alpha_3$ \
\begin{align}
a_1=\alpha_1, \qquad a_2=\alpha_2,\qquad   {a_3} =\frac{2\,\alpha_1+3\,\alpha_3}{8}
\end{align}
equivalently
\begin{align}
\alpha_1={a_1 }, \qquad\alpha_2={a_2},\qquad   \alpha_3=\frac{2}{3}(4\,a_3-a_1).
\end{align}
 This new form yields\footnote{This form agrees with that considered by   Ehlers \cite[Eqs. (46) and (47)]{ehlers2020cosserat}, where the internal energy is 
 	\begin{align}
 	W=&\,\frac{\lambda}{2}\,[\tr(\mathbf{\varepsilon})]^2+\frac{\mu+\mu_{\rm c}}{2} \tr(\mathbf{\varepsilon}\, \mathbf{\varepsilon})+\frac{\mu-\mu_{\rm c}}{2} \tr(\mathbf{\varepsilon}^T \mathbf{\varepsilon})+\frac{\alpha}{2}[\tr(\boldsymbol{\mathfrak{K}})]^2+\frac{\beta+\gamma}{2} \tr(\boldsymbol{\mathfrak{K}}\, \boldsymbol{\mathfrak{K}})+\frac{\beta-\gamma}{2} \tr(\boldsymbol{\mathfrak{K}}^T \boldsymbol{\mathfrak{K}})
 	.\end{align}}
\begin{align}\label{01}
\boldsymbol{\sigma}&:={\rm D}_{ \mathbf{e}} W=2\mu_{\rm e} \, \sym \,\mathbf{e}+2\,\mu_{\rm c} \, \skw\, \mathbf{e}+\lambda_{\rm e} \,\tr (\mathbf{e})\, \id,\notag\\ 
\mathbf{ m}&:={\rm D}_{ \mathbf{\mathfrak{K}}} W =\frac{\mu_{\rm e} L_{\rm c}^2}{2}\,\left[2\,{\alpha_1}\, \sym\,\mathbf{\mathfrak{K}} +2\,{\alpha_2}\,\skw\, \mathbf{\mathfrak{K}}+\alpha_3\,\tr(\mathbf{\mathfrak{K}})\,\id\right].
\end{align}
Using this expression of the internal energy density,   some calculations of constitutive requirements in terms of $\mu_{\rm e},\mu_{\rm c}, \lambda_{\rm e}$ can immediately be extrapolated to similar relations in terms of $\alpha_1, \alpha_2, \alpha_3$, as e.g., the Legendre-Hadamard ellipticity conditions imposed on $W_1(\mathbf{e})$ and $W_2(\mathbf{\mathfrak{K}})$, respectively. 

Anticipating  the importance of having the same structure of these two energy's parts,  let us first treat the generic case of a quadratic form which can then be applied to the balance of linear and angular momentum system. The generic quadratic form we consider is
\begin{align}\label{qf}
\widetilde{W}({\rm D}\phi):=b_1\| \sym \, \text{D} \phi\|^2 +b_2\|\skw \, {\rm D}\phi\|^2 +b_3[{\rm tr}({\rm D}\phi)]^2. \end{align}
Replacing ${\rm D}\phi$ by the rank one dyadic product $\xi \otimes \eta$ we obtain
\begin{align}
{\rm D}^2\widetilde{W}({\rm D}\phi).(\xi\otimes\eta,\xi\otimes\eta)&=b_1\|\sym\xi\otimes\eta\|^2+b_2\|\skw\xi\otimes\eta\|^2+b_3[\tr(\xi\otimes\eta)]^2\notag\\
&=\frac{b_1}{4} \|\xi\otimes\eta+\eta\otimes\xi\|^2+\frac{b_2}{4}\|\xi\otimes\eta-\eta\otimes\xi\|^2+b_3\bigl\langle\xi,\eta\bigr\rangle^2\notag\\
&=\frac{b_1}{4}\|\xi\otimes\eta\|2+2\bigl\langle \xi\otimes\eta,\eta\otimes\xi\bigr\rangle +\frac{b_2}{4} (2\|\xi\otimes\eta\|^2-2\bigl\langle \xi\otimes\eta,\eta\otimes\xi\bigr\rangle) +b_3\bigl\langle \xi,\eta\bigr\rangle^2 \notag\\
&= \frac{b_1}{4}(\|\xi\|^2\|\eta\|^2 +2\bigl\langle \xi,\eta\bigr\rangle^ 2)+ \frac{b_2}{4}(\|\xi\|2\|\eta\|^2-2\bigl\langle \xi,\eta\bigr\rangle 2)+b_3 \bigl\langle \xi,\eta\bigr\rangle^2\\
&= \frac{b_1+b_2}{2} \|\xi\|^2\|\eta\|^2 + \frac{b_1-b_2+2a_3}{2}\bigl\langle \xi,\eta\bigr\rangle^2\notag\\
&= \frac{b_1+b_2}{2} \|\xi\|^2\|\eta\|^2(\sin^2 \theta+\cos^2\theta) + \frac{b_1-b_2+2b_3}{2}\|\xi\|^2\|\eta\|^2 \cos^2\theta \notag\\
&= \frac{b_1+b_2}{2}  \|\xi\|^2\|\eta\|^2 \sin^2\theta+(\frac{b_1+b_2}{2}  +\frac{b_1-b_2+2b_3}{2}\|\xi\|^2\|\eta\|^2 \cos^2\theta\notag\\ 
&= \frac{b_1+b_2}{2} \|\xi\|^2\|\eta\|^2 \sin^2\theta+ (b_1+b_3)\|\xi\|^2\|\eta\|^2 \cos^2\theta, \notag
\end{align}
where $\theta$ represents the angle between $\xi $ and $\eta$.

Thus, the bilinear form of the second derivative of $\widetilde{W}$ reads
\begin{align}
{\rm D}^2\widetilde{W}({\rm D}\phi).(\xi\otimes\eta,\xi\otimes\eta)=(b_1 +b_2)\|\xi\|^2\|\eta\|^2 \sin^2\theta+(b_1 +b_3)\|\xi\|^2\|\eta\|^2\cos^2\theta,
\end{align}
and  we infer the necessary and sufficient conditions for strict Legendre-Hadamard ellipticity of the quadratic form $\widetilde{W}$, i.e., 
\begin{align}
{\rm D}^2\widetilde{W}({\rm D}\phi).(\xi\otimes\eta,\xi\otimes\eta)>0\qquad \forall\  \eta, \xi\in \mathbb{R}^3, \ \ \lVert \eta\rVert=\lVert \xi\rVert=1,
\end{align}
to be 
\begin{align}
b_1 +b_2>0, \qquad b_1 + b_3 >0. 
\end{align} 
Applying this result to both the strain energy $W_1(\mathbf{e})$  and the curvature energy $W_2(\boldsymbol{\mathfrak{K}})$, see \eqref{enersplit},   we obtain the Legendre-Hadamard ellipticity conditions for  both energy's part, see Subsection \ref{LHs}
\begin{align}\label{dse12}
 \mu_{\rm e} +\mu_{\rm c} >0, \qquad \qquad 2\mu_{\rm e} +\lambda_{\rm e}  >0,
\end{align}
and formally similar
\begin{align} \label{dse13} \alpha_1+\alpha_2>0,\qquad \qquad 2\,\alpha_1+\alpha_3>0,
\end{align}
respectively.

 Using the weight parameters $\alpha_1, \alpha_2, \alpha_3$, the form of the system of partial differential equations is
\begin{align}\label{PDE11}
\rho\,\,u_{i,tt}&=(\mu_{\rm e} +\mu_{\rm c})\dd   u_{i,jj}+(\mu_{\rm e} -\mu_{\rm c}+\lambda_{\rm e}  )\dd  u_{j,ij}+2\,\mu_{\rm c} \,  \epsilon _{ijk}\,\vartheta _{k,j},
\vspace{1.2mm}\\
2\,\rho\, \eta\,\mu_{\rm e} \,\tau_{\rm c}^2\,\vartheta_{i,tt}&=\frac{\mu_{\rm e} {L_{\rm c}^2}}{2}\,\left[(\alpha_1+\alpha_2)\dd  \vartheta _{i,jj} +(\alpha_1-\alpha_2+\alpha_3)\vartheta _{j,ij}\right]+2\,\mu_{\rm c} \,  \varepsilon _{ijk}\,u_{k,j}-4\, \mu_{\rm c}\, \vartheta _i \, . \notag
\end{align}

\section{Identification of parameters in the dislocation formulation}\label{idsec}\setcounter{equation}{0}

We immediately remark that $\mathbf{e}={\mathbf{e}^*}^T$ while $\boldsymbol{\alpha}$ is not directly related to the curvature stress tensor $\boldsymbol{\mathfrak{K}}$ considered by Eringen.  However, due to the   Curl-$\mathrm{D}\,\axl\,$ converting formulae \eqref{curlaxl}, the internal energy density proposed here admits the alternative expression 
\begin{align}
W=\, \mu_{\rm e} &\|\sym \,(\mathrm{D}u -\mathbf{A})\|^2+\mu_{\rm c} \|\skw(\mathrm{D}u -\mathbf{A})\|^2+ \frac{\lambda_{\rm e} }{2}\, [\tr(\mathrm{D}u -\mathbf{A})]^2\notag\\&
  +\frac{\mu_{\rm e} L_{\rm c}^2}{2}\left[{a_1}\|\dev  \,\sym \,\Curl\, \mathbf{A}\|^2 +{a_2}\| \skw \,\Curl\, \mathbf{A}\|+ \frac{a_3}{3}\, \tr(\Curl\, \mathbf{A})^2\right]\notag\\=\,\mu_{\rm e} &\,\lVert\dev   \, \sym\,\mathbf{e}^* \rVert ^{2}+\mu_{\rm c}\,\lVert \skw\,\mathbf{e}^* \rVert ^{2}+\frac{2\mu_{\rm e} +3\,\lambda_{\rm e}  }{6}\left[\mathrm{tr} \left(\mathbf{e}^* \right)\right]^{2}\\& +\frac{\mu_{\rm e} L_{\rm c}^2}{2}\,\left[a_1\,\lVert\dev   \,\sym\,\boldsymbol{\mathfrak{K}} \rVert ^{2}+a_2\,\lVert \skw\,\boldsymbol{\mathfrak{K}} \rVert ^{2}+\frac{4\,a_3}{3}\left[\mathrm{tr} (\boldsymbol{\mathfrak{K}} )\right]^{2}\right].\notag
\end{align}
  Due to the  orthogonal Cartan-decomposition of the Lie-algebra 
  \begin{align}
\mathfrak{gl}(3)=\{\mathfrak{sl}(3)\cap \Sym(3)\}\oplus\mathfrak{so}(3)\oplus\mathbb{R}\cdot\id,\qquad 
\mathbf{X}=\dev  \,\sym\, \mathbf{X}+\skw\, \mathbf{X}+\frac{1}{3}\tr( \mathbf{X})\cdot\id\
\end{align}
and the identity
\begin{align}
\bigl\langle \mathbf{X}, \mathbf{Y}\bigr\rangle= \bigl\langle\dev\,\sym\, \mathbf{X},\dev  \,\sym\,\mathbf{Y}\bigr\rangle+\bigl\langle\skw\, \mathbf{X}, \skw\,\mathbf{Y}\bigr\rangle+\frac{1}{3}\tr( \mathbf{X})\tr( \mathbf{Y})
\end{align}
 the expression of the internal energy  density proposed by Eringen may be written as
\begin{align}
W= \frac{1}{2}\Big[&(\mu^*+\varkappa)  \Big(\bigl\langle\dev\,\sym\, \mathbf{e}^* ,\dev  \,\sym\,\mathbf{e}^* \bigr\rangle+\bigl\langle\skw\, {\rm e}^* , \skw\,\mathbf{e}^* \bigr\rangle+\frac{1}{3}\tr( \mathbf{e}^* )\tr( \mathbf{e}^* ) \Big)\notag\\&+ \mu^*\Big(\bigl\langle\dev\,\sym\, \mathbf{e}^* ,\dev  \,\sym\,\mathbf{e}^* \bigr\rangle-\bigl\langle\skw\, \mathbf{e}^* , \skw\,\mathbf{e}^* \bigr\rangle+\frac{1}{3}\tr( \mathbf{e}^* )\tr( \mathbf{e}^* ))\Big)+\lambda_{\rm e} \left(\tr \mathbf{e}^* \right)^2\\&+\gamma  \Big(\bigl\langle\dev\,\sym\, \boldsymbol{\mathfrak{K}} ,\dev  \,\sym\,\boldsymbol{\mathfrak{K}} \bigr\rangle+\bigl\langle\skw\, \boldsymbol{\mathfrak{K}} , \skw\,\boldsymbol{\mathfrak{K}} \bigr\rangle+\frac{1}{3}\tr( \boldsymbol{\mathfrak{K}} )\tr( \boldsymbol{\mathfrak{K}} ) \Big)\notag\\&+\beta  \Big(\bigl\langle\dev\,\sym\, \boldsymbol{\mathfrak{K}} ,\dev  \,\sym\,\boldsymbol{\mathfrak{K}} \bigr\rangle-\bigl\langle\skw\, \boldsymbol{\mathfrak{K}} , \skw\,\boldsymbol{\mathfrak{K}} \bigr\rangle+\frac{1}{3}\tr( \boldsymbol{\mathfrak{K}} )\tr( \boldsymbol{\mathfrak{K}} ) \Big)+ \alpha  \left(\tr \boldsymbol{\mathfrak{K}} \right)^2\Big]\notag\\
=\frac{1}{2}\Big[&(2\,\mu^*+\varkappa)  \|\dev\,\sym\, \mathbf{e}^* \|^2+\varkappa\, \|\skw\, \mathbf{e}^* \|^2+\frac{2\,\mu^*+\varkappa+3\,\lambda}{3}[\tr( \mathbf{e}^* )]^2\notag\\&+(\gamma+\beta ) \|\dev\,\sym\, \boldsymbol{\mathfrak{K}} \|^2+(\gamma-\beta ) \|\skw\, \boldsymbol{\mathfrak{K}} \|^2+\frac{\beta +\gamma +3\,\alpha }{3}[\tr( \boldsymbol{\mathfrak{K}} )]^2\Big].\notag
\end{align}

Using again the  orthogonal Cartan-decomposition of the Lie-algebra $
\mathfrak{gl}(3)
$ and since $\mathbf{e}^*$ and $\boldsymbol{\mathfrak{K}} $ are independent constitutive variables, we obtain the  relations between the parameters in the two forms 
\begin{align}
\mu_{\rm e} &=\mu^*+\frac{\varkappa}{2},\qquad\ \   \mu_{\rm c}=\frac{\varkappa}{2}\,, \qquad \qquad \ \ \ \frac{2\mu_{\rm e} +3\,\lambda  }{6}=\frac{2\,\mu^*+\varkappa+3\,\lambda}{6},\notag\\
{\mu_{\rm e} L_{\rm c}^2}\, {a_1 }&=\gamma+\beta , \qquad {\mu_{\rm e} L_{\rm c}^2}\,{a_2}=\gamma-\beta , \qquad {\mu_{\rm e} L_{\rm c}^2}\,\frac{4\,{a_3}}{3}=\frac{\beta +\gamma +3\,\alpha }{3},\qquad 
2\, \eta\,\mu_{\rm e} \,\tau_{\rm c}^2=j,
\end{align}
which implies
\begin{align}
\lambda_{\text{e}} &=\lambda \, ,\qquad 
\mu_{\text{e}} =\mu^* +\frac{\varkappa}{2}\, ,\qquad 
\mu_{\text{c}} =\frac{\varkappa}{2}\, ,\notag\\
a_1 &=\frac{ \gamma+\beta    }{L_{\rm c}^2 \, (\mu^{*} +\frac{\varkappa}{2})} \, ,\qquad 
a_2 =\frac{\gamma-\beta  }{L_{\rm c}^2 \, (\mu^{*} +\frac{\varkappa}{2})}\, ,\qquad 
a_3 =\frac{3\alpha  + \beta  +\gamma }{4\,L_{\rm c}^2 \, (\mu^{*} +\frac{\varkappa}{2})} \, ,\qquad 
\eta=\frac{j}{2\, (\mu^{*} +\frac{\varkappa}{2})\,\tau_{\rm c}^2},
\end{align}
or equivalently,  expressed in the weight parameters $\alpha_1, \alpha_2, \alpha_3$
\begin{align}
\lambda_{\text{e}} &=\lambda \, ,\qquad 
\mu_{\text{e}} =\mu^* +\frac{\varkappa}{2}\, ,\qquad 
\mu_{\text{c}} =\frac{\varkappa}{2}\, ,\notag\\
\alpha_1&=\frac{ \gamma+\beta   }{L_{\rm c}^2 \, (\mu^{*} +\frac{\varkappa}{2})} \, ,\qquad 
\alpha_2=\frac{\gamma-\beta  }{L_{\rm c}^2 \, (\mu^{*} +\frac{\varkappa}{2})}\, ,\qquad 
\alpha_3 =\frac{2\,\alpha}{L_{\rm c}^2 \, (\mu^{*} +\frac{\varkappa}{2})} \, ,\qquad
\eta=\frac{j}{2\, (\mu^{*} +\frac{\varkappa}{2})\,\tau_{\rm c}^2},
\end{align}

And in the other way
\begin{align}
\lambda &=\lambda_{\text{e}} \, ,\qquad 
\mu^* =\mu_{\text{e}} - \mu_{\text{c}}\, ,\qquad 
\varkappa  =2\,\mu_{\text{c}}\, ,\notag\\
\gamma  &= \frac{a_1+a_2}{2} \, L_{\rm c}^2 \,  \,\mu_{\text{e}}=\frac{\alpha_1+\alpha_2}{2} \, L_{\rm c}^2 \,  \,\mu_{\text{e}} \,,\quad 
\beta  =\frac{a_1-a_2}{2} \, L_{\rm c}^2 \,  \,\mu_{\text{e}}= \frac{\alpha_1-\alpha_2}{2} \, L_{\rm c}^2 \,  \,\mu_{\text{e}}\,, \\ \alpha  &=\frac{1}{3}L_{\rm c}^2 \,  \,\mu_{\text{e}} (4\, a_3-a_1)= \frac{1}{2}\,L_{\rm c}^2 \, \mu_{\text{e}}\alpha_3\,,\qquad 
j=2\, \eta\,\mu_{\rm e} \,\tau_{\rm c}^2\,.\notag
\end{align}

\section{Constitutive assumptions  in the new parameter set}\setcounter{equation}{0}\label{Const}

\subsection{Positive definiteness  of the elastic energy density and well-posedness of the solution\label{EN}}

Using the following equivalent alternative forms of the internal energy density, namely \begin{align}
W=& \,\mu_{\rm e} \,\lVert\dev   \, \sym\,\mathbf{e}\rVert ^{2}+\mu_{\rm c}\,\lVert \skw\,\mathbf{e}\rVert ^{2}+\frac{2\mu_{\rm e} +3\,\lambda_{\rm e}  }{6}\left[\mathrm{tr} \left(\mathbf{e}\right)\right]^{2}\notag\\&
\quad \quad  +\frac{\mu_{\rm e} L_{\rm c}^2}{2}\left[{a_1}\|\dev  \,\sym \,\Curl\, \mathbf{A}\|^2 +{a_2}\| \skw \,\Curl\, \mathbf{A}\|+ \frac{a_3}{3}\, \tr(\Curl\, \mathbf{A})^2\right]\\
=&\,\mu_{\rm e} \,\lVert\dev   \, \sym\,\mathbf{e}\rVert ^{2}+\mu_{\rm c}\,\lVert \skw\,\mathbf{e}\rVert ^{2}+\frac{2\mu_{\rm e} +3\,\lambda_{\rm e}  }{6}\left[\mathrm{tr} \left(\mathbf{e}\right)\right]^{2}\notag\\&
\quad \quad  +\frac{\mu_{\rm e} L_{\rm c}^2}{2}\left[{\alpha_1}\|\dev  \,\sym \,\Curl\, \mathbf{A}\|^2 +{\alpha_2}\| \skw \,\Curl\, \mathbf{A}\|+ \frac{2\,\alpha_1+3\,\alpha_3}{6}\, \tr(\Curl\, \mathbf{A})^2\right]\notag\\=&\,\mu_{\rm e} \,\lVert\dev   \, \sym\,\mathbf{e}^* \rVert ^{2}+\mu_{\rm c}\,\lVert \skw\,\mathbf{e}^* \rVert ^{2}+\frac{2\mu_{\rm e} +3\,\lambda_{\rm e}  }{6}\left[\mathrm{tr} \left(\mathbf{e}^* \right)\right]^{2}\notag\\& +\frac{\mu_{\rm e} L_{\rm c}^2}{2}\,\left[\ a_1 \,\lVert\dev   \,\sym\,\boldsymbol{\mathfrak{K}} \rVert ^{2}+a_2\,\lVert \skw\,\boldsymbol{\mathfrak{K}} \rVert ^{2}+\frac{ 4\,{a_3}}{3}\left[\mathrm{tr} \left(\boldsymbol{\mathfrak{K}} \right)\right]^{2}\right]\\
=&\, \mu_{\rm e} \,\lVert\dev   \, \sym\,\mathbf{e}\rVert ^{2}+\mu_{\rm c}\,\lVert \skw\,\mathbf{e}\rVert ^{2}+\frac{2\mu_{\rm e} +3\,\lambda_{\rm e} }{6}\left[\mathrm{tr} \left(\mathbf{e}\right)\right]^{2}\notag\\& +\frac{\mu_{\rm e} L_{\rm c}^2}{2}\,\left[\ {\alpha_1}\,\lVert\dev   \,\sym\,\mathbf{\mathfrak{K}}\rVert ^{2}+{\alpha_2}\,\lVert \skw\,\mathbf{\mathfrak{K}}\rVert ^{2}+\frac{2\,\alpha_1+3\,\alpha_3}{6}\left[\mathrm{tr} \left(\mathbf{\mathfrak{K}}\right)\right]^{2}\right],\notag
\end{align}
and due to the orthogonal Cartan-decomposition of the Lie-algebra $\mathfrak{gl}(3)$,	
 the  strict positive definiteness as function of the strain tensors $\mathbf{e}={\rm D}u-\mathbf{A} $ and $\boldsymbol{\alpha}=-\Curl\, \mathbf{A}$, as well as function of the strain tensors $\mathbf{e}^*=({\rm D}u)^T+\mathbf{A}$ and $\mathfrak{K}={\rm D} \vartheta={\rm D} (\axl \mathbf{A})$,  is equivalent to 
\begin{align}\label{posCoss2}
\quad\mu_{\rm e}  &>0, \qquad \quad\mu_{\rm c}>0,\qquad \quad  2\mu_{\rm e} +3\,\lambda_{\rm e}  >0,\\
a_1&>0, \qquad \quad a_2>0, \qquad\qquad\quad \quad\, \, a_3>0. \notag
\end{align}

Moreover, \eqref{posCoss2}$_2$ is equivalent to
\begin{align}\label{pden}
\alpha_1> 0, \qquad \quad\alpha_2> 0, \qquad \quad 2\,\alpha_1+3\,\alpha_3> 0.
\end{align}

Note that in the linear Cosserat model it is not possible to set the Cosserat couple modulus $\mu_{\rm c}=0$ since otherwise the constitutive coupling of the fields is lost (a coupling may remain due to some non-standard boundary conditions, but this is a pathological case \cite{Lakes85}).

Under the assumption of positive definiteness, it was proven that in both  the dynamic and the static case, and for both formulation, in terms of $\Curl \mathbf{A}$ and ${\rm D} (\axl \mathbf{A})$, the obtained mathematical model has a unique solution. 

However, due to some new Korn-type inequality \cite{Neff_JeongMMS08,lewintan2021korn,lewintan2021lp,lewintan2021nevcas}, the  existence and uniqueness of the solution is still valid  for 
\begin{align}\
\quad\mu_{\rm e}  > 0, \qquad \quad\mu_{\rm c}> 0,\qquad \quad  2\mu_{\rm e} +3\,\lambda_{\rm e}  > 0,\qquad \quad \alpha_1> 0, \qquad \quad\alpha_2\geq 0, \qquad \quad 2\,\alpha_1+3\,\alpha_3\geq 0 \, ,
 \label{posCoss1}
\end{align}

or equivalently
\begin{align}
	\quad\mu_{\rm e}  > 0, \qquad \quad\mu_{\rm c}> 0,\qquad \quad  2\mu_{\rm e} +3\,\lambda_{\rm e}  > 0,\qquad \quad a_1> 0, \qquad \quad a_2\geq 0, \qquad \quad a_3\geq 0 \, ,
	\label{posCoss1b}
\end{align}
i.e., under weaker assumptions on the constitutive parameters, in comparison to the positive definiteness conditions of the internal energy density, see the last two inequalities which allow $\alpha_2$ and $2\,\alpha_1+3\,\alpha_3$ to vanish and the curvature energy may be chosen as  (\textbf{the conformal curvature case})
\begin{align}
\frac{\mu_{\rm e} L_{\rm c}^2}{2}{\alpha_1}&\|\dev  \,\sym \,\Curl\, \mathbf{A}\|^2=\frac{\mu_{\rm e} L_{\rm c}^2}{2}{a_1}\|\dev  \,\sym \,\Curl\, \mathbf{A}\|^2  \notag\\&=\frac{\mu_{\rm e} L_{\rm c}^2}{2}\, a_1 \,\lVert\dev   \,\sym\,\boldsymbol{\mathfrak{K}} \rVert ^{2}=\frac{\mu_{\rm e} L_{\rm c}^2}{2}\, {\alpha_1}\,\lVert\dev   \,\sym\,\boldsymbol{\mathfrak{K}}\rVert ^{2}=\frac{\mu_{\rm e} L_{\rm c}^2}{2}\, {\alpha_1}\,\lVert\dev   \,\sym\,{\rm D}\vartheta\rVert ^{2}.
\end{align}

This choice is mandatory for stable identification of Cosserat parameters in the sense discussed in \cite{neff2010stable}.

\subsection{Real plane waves in isotropic Cosserat elastic solids}\label{Rpw}

We say that there exists  {\it real plane waves}  in the  direction $\xi=(\xi_1,\xi_2,\xi_3)$, $\lVert{\xi}\rVert^2=1$,  if for every  wave number $k>0$  the system of partial differential equations \eqref{PDE} admits a solution in the form:
\begin{align}\label{ansatzwp}
u(x_1,x_2,x_3,t)&=\underbrace{\begin{footnotesize}\begin{pmatrix}\widehat{u}_1\\\widehat{u}_2\\\widehat{u}_3\end{pmatrix}\end{footnotesize}}_{:=\,\widehat{u}}
\, e^{{\rm i}\, \left(k\langle {\xi},\, x\rangle_{\mathbb{R}^3}-\,\omega \,t\right)}\,,\\ \vartheta (x_1,x_2,x_3,t)&=\underbrace{\begin{footnotesize}{\rm i}\,\begin{pmatrix}\widehat{\vartheta }_1\\\widehat{\vartheta }_2\\\widehat{\vartheta }_3\end{pmatrix}\end{footnotesize}}_{:=\,\widehat{\vartheta}} \, e^{{\rm i}\, \left(k\langle {\xi},\, x\rangle_{\mathbb{R}^3}-\,\omega \,t\right)},\quad 
\widehat{u}, \widehat{ \vartheta   }\in\mathbb{C}^{3}, \quad (\widehat{u}, \widehat{ \vartheta   })^T\neq 0\,,\notag
\end{align}
only for  real frequencies $\omega\in \mathbb{R}$, where ${\rm i}\,=\sqrt{-1}$ is the complex unit.  The plane wave is called ``real'' since it is defined by real values of $\omega$. Note that we take ${\rm i}\,\widehat{\vartheta }$  since this choice will lead us in the end to only  real valued matrices. Otherwise, we would have to deal with complex valued matrices in the linear Cosserat theory. 

There exist real plane waves if for every  wave number $k>0$ the following  systems of  equations \eqref{PDE} admit  non-trivial solutions:   
\begin{align}\label{fpm}
[\mathbf{\mathbf{Q}}_1(e_1,k)-\omega^2\widehat{\id}]\, w&=0 \qquad 
w=\begin{footnotesize}\begin{pmatrix}
\widehat{u}_1,\widehat{u}_2,\widehat{ \vartheta }_3
\end{pmatrix}\end{footnotesize}^T, \\
[\mathbf{\mathbf{Q}}_2(e_1,k)-\omega^2\widetilde{\id} ]\, w&=0 \qquad 
w=\begin{footnotesize}\begin{pmatrix}
\widehat{u}_3,\widehat{ \vartheta }_1,\widehat{ \vartheta }_2
\end{pmatrix}\end{footnotesize}^T\notag
\end{align}
only for real frequencies $\omega\in \mathbb{R}$,
where
\begin{align}
\notag\mathbf{\mathbf{Q}}_1(e_1,k)&=\begin{footnotesize}\begin{footnotesize}\begin{pmatrix}
k^2(2\, \mu_{\rm e} +\lambda_{\rm e}  )& 0& 0\vspace{2mm}\\
0& k^2(\mu_{\rm e} +\mu_{\rm c})& -2\, k\, \mu_{\rm c}\vspace{2mm}\\
0& -2\, k\, \mu_{\rm c}&k^2\, {\mu_{\rm e} \,L_{\rm c}^2}\,(\alpha_1 +\alpha_2)+4\, \mu_{\rm c}
\end{pmatrix}\end{footnotesize}\end{footnotesize},\\
\mathbf{\mathbf{Q}}_2(e_1,k)&=\begin{footnotesize}\begin{footnotesize}\begin{pmatrix}
k^2(\mu_{\rm e} +\mu_{\rm c})& 0& 2\, k\, \mu_{\rm c}\vspace{2mm}\\
0&k^2\,{\mu_{\rm e} \,L_{\rm c}^2}\,(2\,\alpha_1+\alpha_3)+4 \mu_{\rm c}& 0\vspace{2mm}\\
2\, k\, \mu_{\rm c}& 0&k^2\, {\mu_{\rm e} \,L_{\rm c}^2}\,(\alpha_1 +\alpha_2)+4\, \mu_{\rm c}
\end{pmatrix}\end{footnotesize}\end{footnotesize},\\
\widehat{\id}&=\begin{footnotesize}\begin{pmatrix} \rho&0	&0
\\
0&\rho 
&0
\\0
& 0 & \rho\,j\,\mu_{\rm e} \,\tau_{\rm c}^2\, \end{pmatrix}\end{footnotesize}, \qquad \qquad 
\widetilde{\id} =\begin{footnotesize}\begin{pmatrix} \rho&0	&0
\\
0&\rho\,j\,\mu_{\rm e} \,\tau_{\rm c}^2
&0
\\0
& 0 & \rho\,j\,\mu_{\rm e} \,\tau_{\rm c}^2\end{pmatrix}\end{footnotesize}.\notag
\end{align}

Then, for $k>0$ and due to the isotropy, extrapolating to all directions of propagation \cite{neff2017real}, we have
\begin{proposition}\label{proprealw} The necessary and sufficient conditions for existence of  real planar waves in any direction ${\xi}\in \mathbb{R}^3$, $\xi\neq 0$, in the framework of the linear isotropic elastic Cosserat theory \cite{neff2017real,khan2022existence} are
	\begin{align}\label{d11}
	2\mu_{\rm e} +\lambda_{\rm e}  >0,\qquad \qquad \mu_{\rm e} >0,\qquad \qquad \mu_{\rm c} >0, \qquad \qquad 2\,\alpha_1+\alpha_3>0,\qquad \qquad   \alpha_1+\alpha_2>0,
	\end{align}
	equivalently
		\begin{align}\label{d1a}
	2\mu_{\rm e} +\lambda_{\rm e}  >0,\qquad \qquad \mu_{\rm e} >0,\qquad \qquad \mu_{\rm c} >0, \qquad \qquad a_1+2\,a_3>0,\qquad \qquad   a_1+a_2>0.
	\end{align}
\end{proposition}

\bigskip
	\begin{table}[h!]\begin{center}
			\begin{tabular}{ |c| c | c |}
				\hline 
				Name&	Expression& \begin{minipage}{3.5cm} 
					Dispersive waves/\\Non-dispersive waves	\end{minipage}\\
				\hline 
				\begin{minipage}{7cm}\medskip
					the velocity of the acoustic branch of translational
					compression (longitudinal) plane wave \medskip
				\end{minipage}  & $\mathfrak{c}_p= \sqrt{\frac{\lambda_e+2\mu_e
					}{\rho}}$ &  non-dispersive \\\hline 
				\begin{minipage}{7cm}\medskip
					the limit of the group/phase velocity of the acoustic
					branch of the shear–rotational wave at $\omega\to 0$ ($k\to 0$)  \medskip
				\end{minipage} & $\mathfrak{c}_{t} = \sqrt{\frac{\mu_e}{\rho}}$&  dispersive 
				\\\hline 
				\begin{minipage}{7cm}\medskip
					the limit of the group/phase velocity of the acoustic
					branch of the shear–rotational wave at $\omega\to \infty$ ($k\to \infty$)\medskip
				\end{minipage} & $\mathfrak{c}_s=\sqrt{\frac{\mu_e+\mu_c}{\rho}}$ &  dispersive  
				\\
				\hline 
				\begin{minipage}{7cm}\medskip
					the group/phase velocity for the compressional rotational wave in the limit $\omega\to \infty$ ($k\to \infty$)\medskip
				\end{minipage} & $\mathfrak{c}_{m,p}=\sqrt{\frac{L_c^2(2\alpha_1+\alpha_3)}{\rho \, j\, \tau_c^2}}$ &  dispersive 
				\\\hline 
				\begin{minipage}{7cm}\medskip
					the limit  of the group/phase velocity of the acoustic
					branch of the shear–rotational wave  at $\omega\to \infty$ ($k\to\infty$)\medskip
				\end{minipage} & $\mathfrak{c}_{m,s}=\sqrt{\frac{L_c^2(\alpha_1+\alpha_2)}{\rho \, j\, \tau_c^2}}=\sqrt{\frac{L_c^2\,\gamma}{\rho \, j\, \tau_c^2}}$ &  dispersive \\\hline 
				\begin{minipage}{7cm}\medskip
					the limit of the optical branch (compressional-rotational and shear-rotational)  at the cut-off frequency $\omega=2\sqrt{\frac{\mu_c}{\rho j \mu_e \tau_c^2}}$, $k=0$\medskip
				\end{minipage} & \begin{minipage}{3cm}$0$ (group velocity)\\ $/\infty$ (phase
					velocity) \end{minipage}&  dispersive  \\\hline
			\end{tabular}
			\caption{Group velocities $c=\frac{\omega}{k}$ or/and phase velocities   $\frac{d\omega}{dk}$ and the cut-off frequency in linear Cosserat elasticity, see \cite{khan2022existence}.}\label{T1}\end{center} 
\end{table}

\begin{remark}\label{iterpretation1}
	Using the notation from Table \ref{T1}, we have the following  interpretation:
		\begin{itemize}\item[i)] the first implication of the set of conditions
			\eqref{d11} means that all these waves (compressional/shear-rotational  waves, acoustic/optical branch) are real;
			Once this aspect  is clarified, we can treat and interpret further the propagation of plane waves;
			\item[ii)] under conditions 	\eqref{d11} all branches of waves are real for the entire range $[0,\infty)$ of the frequency;
			\item[iii)] we  can also see directly from  the first condition that the translational compressional wave
			is real;
			\item[iv)]  the second means that the acoustic branch of shear–rotational
			wave is real at low frequencies and together with the third means that  the acoustic branch of shear–rotational
			wave is real at high frequencies;
			\item[v)]  the fourth implies that the optical branch of the
			shear–rotational wave is real at high frequencies;
			\item[vi)]  the third one means that the
			optical branch of the shear–rotational wave at high frequencies has a larger
			velocity than the acoustic branch of the same wave at low frequencies (if they
			both exist, which is the case due to other conditions); 
			\item[vii)] 	the fifth one expresses directly that   the compressional rotational
			wave at high frequencies is real.
	\end{itemize}
\end{remark}

\subsection{Strong ellipticity (rank-one convexity, Legendre-Hadamard ellipticity) \\ conditions}\label{LHs}

		In  linear isotropic classical elasticity, the necessary and sufficient conditions for existence of  real planar waves in any direction ${\xi}\in \mathbb{R}^3$, $\xi\neq 0$ are
		\begin{align}\label{d1ce}
		2\mu_{\rm e} +\lambda_{\rm e}  >0,\qquad \qquad \mu_{\rm e} >0,
		\end{align}
		and they are equivalent to the strong ellipticity conditions (Legendre-Hadamard ellipticity).

		 For the Cosserat (micropolar) model  the necessary and sufficient conditions for existence of a real planar wave are slightly different compared to  the    strong ellipticity conditions  (Legendre–Hadamard ellipticity), the    strong ellipticity conditions being connected to acceleration waves.   In our notation,
		the strong
		ellipticity condition for  Cosserat media is represented by
		the inequality \cite{eremeyev2007constitutive,Eremeyev4,shirani2020legendre}
		\begin{align}
		\frac{d^2\, }{d\, \tau^2}W(\mathbf{e} + \tau \, \xi \otimes \eta, \boldsymbol{\mathfrak{K}} + \tau \, \zeta\otimes \eta)\bigg|_{\tau=0}>0 \qquad \forall\  \eta, \xi, \zeta \in \mathbb{R}^3, \ \ \lVert \eta\rVert=\lVert \xi\rVert=\lVert \zeta\rVert=1.
		\end{align}

	The Legendre–Hadamard ellipticity condition \cite{eremeyev2007constitutive,Eremeyev4,shirani2020legendre} is equivalent to the positive definiteness of the  {\it acoustic tensor} 
\begin{align}
\widehat{\mathbf{Q}}=\begin{footnotesize}\begin{pmatrix}
\widehat{\mathbf{Q}}_1&0\\0&\widehat{\mathbf{Q}}_2
\end{pmatrix}\end{footnotesize}
\end{align}
whose block matrices are defined through
\begin{align}
\widehat{\mathbf{Q}}_1(\xi)
=&\frac{1}{2}\begin{footnotesize}\begin{footnotesize}\begin{pmatrix}
(2\mu_{\rm e} +\lambda_{\rm e}  )\dd\xi_1^2+(\mu_{\rm e} +\mu_{\rm c})(\xi_2^2+\xi_3^2)& (\mu_{\rm e} -\mu_{\rm c}+\lambda_{\rm e}  )\xi_1\xi_2 &(\mu_{\rm e} -\mu_{\rm c}+\lambda_{\rm e}  )\xi_1\xi_3\vspace{2mm}\\
(\mu_{\rm e} -\mu_{\rm c}+\lambda_{\rm e}  )\xi_1\xi_2& 	(2\mu_{\rm e} +\lambda_{\rm e}  )\dd\xi_2^2+(\mu_{\rm e} +\mu_{\rm c})(\xi_3^2+\xi_1^2)&(\mu_{\rm e} -\mu_{\rm c}+\lambda_{\rm e}  )\xi_2\xi_3\vspace{2mm}\\
(\mu_{\rm e} -\mu_{\rm c}+\lambda_{\rm e}  )\xi_1\xi_3& (\mu_{\rm e} -\mu_{\rm c}+\lambda_{\rm e}  )\xi_2\xi_3&	(2\mu_{\rm e} +\lambda_{\rm e}  )\dd\xi_3^2+(\mu_{\rm e} +\mu_{\rm c})(\xi_1^2+\xi_2^2)
\end{pmatrix}\end{footnotesize}\end{footnotesize}\notag
\end{align}
and
\begin{align}
\widehat{\mathbf{Q}}_2(\zeta)
=&\frac{1}{2}\begin{footnotesize}\begin{pmatrix}
(2\,\alpha_1 +\alpha_3)\zeta_1^2+(\alpha_1+\alpha_2 )(\zeta_1^2+\zeta_3^2)& (\alpha_1 -\alpha_2+\alpha_3)\zeta_1\zeta_2 &(\alpha_1 -\alpha_2+\alpha_3 )\zeta_1\zeta_3\\
(\alpha_1 -\alpha_2+\alpha_3 )\zeta_1\zeta_2& (2\,\alpha_1 +\alpha_3)\zeta_2^2+(\alpha_1+\alpha_2 )(\zeta_3^2+\zeta_1^2)&(\alpha_1 -\alpha_2+\alpha_3 )\zeta_2\zeta_3\\
(\alpha_1 -\alpha_2+\alpha_3)\zeta_1\zeta_3& (\alpha_1-\alpha_2+\alpha_3 )\zeta_2\zeta_3&(2\,\alpha_1 +\alpha_3)\zeta_3^2+(\alpha_1+\alpha_2 )(\zeta_1^2+\zeta_2^2)
\end{pmatrix}\end{footnotesize},\notag
\end{align}
for any
nonzero wave directions $\xi\in \mathbb{R}^3$ and $\zeta\in \mathbb{R}^3$.

The positive definiteness of the acoustic tensor, i.e., the Legendre-Hadamard ellipticity conditions, is studied in  \cite{shirani2020legendre,eremeyev2007constitutive,Eremeyev4}  and, see \eqref{dse12} and \eqref{dse13}, it is satisfied if and only if
\begin{align}\label{dse1}
2\,\mu_{\rm e} +\lambda_{\rm e}  >0,\qquad \quad \mu_{\rm e}+\mu_{\rm c} >0,\qquad \quad   \quad 2\,\alpha_1+\alpha_3>0, \qquad \alpha_1+\alpha_2>0,
\end{align}
equivalently
\begin{align}\label{d1a}
2\,\mu_{\rm e} +\lambda_{\rm e}  >0,\qquad \qquad \mu_{\rm e} +\mu_{\rm c} >0, \qquad \qquad a_1+2\, a_3>0,\qquad \qquad   a_1+a_2>0.
\end{align}
  The absence of a coupling between
$\mathbf{e}$ and $\boldsymbol{\mathfrak{K}}$ in the strain energy  leads to a simplification of the calculations. 
The conditions   \eqref{dse1} (strong ellipticity conditions, Legendre-Hadamard ellipticity, the positive definiteness of the acoustic tensor) imply  the existence of  real translational compressional
waves in the entire range of real frequencies, of  real shear rotational waves (both branches) at high frequencies, and of real
rotational compressional wave at high frequencies, but at lower frequencies the
latter waves may not be real since \eqref{dse1} does not guaranty that $\mathfrak{c}_t=\frac{\mu_{\rm e} }{\rho}$ is real, since the positivity of $\mu_{\rm e}$ is not necessarily implied by \eqref{dse1}. To the contrary, the conditions \eqref{d11} imply that all these branches and types of plane wave are real, i.e., the group/phase velocities are real on the entire range of possible frequencies.

The strong ellipticity conditions \eqref{dse1} are weaker than the conditions \eqref{d11} in the sense that they are implied by the necessary and sufficient conditions for existence of a real planar wave (i.e., they imply the strong ellipticity and, therefore, the considered PDEs system is not unstable)  but not vice versa. However, the strong ellipticity conditions \eqref{dse1} are not sufficient for some applications,   and we believe that they are also not suitable for any approach regarding the propagation of  Rayleigh waves in Cosserat solids \cite{khan2022existence}.  Of course positive definiteness or the weaker conditions \eqref{posCoss1} of the elastic energy should be sufficient for any application.


\section{Further identifications for micropolar constants}\setcounter{equation}{0}\label{fnot}

\subsection{Eringen's initial notation: the notation used by Lakes }

The first notations and tensors used by Eringen \cite{Eringen68} and then in the Lakes group's  experiments \cite{Lakes85,lakes2016physical} are
\begin{align}\label{eq:stress_classic}
\widetilde{\sigma}_{ji} &=
(2\,\mu^*+\varkappa)  ( u_{i,j}+u_{j,i})+\varkappa \, \epsilon_{jim}\Big(\underbrace{\frac{1}{2}\epsilon_{mks}u_{s,k}}_{r_m}-\vartheta_m\Big)+\lambda \, u_{k,k}\delta_{ji},
\\*
m _{ji}&=\beta \,\vartheta_{j,i}+\gamma\,
 \vartheta_{i,j}+ \alpha  \, \vartheta_{k,k}\delta_{ji} 
\, .
\notag
\end{align}
The vector  $r_m:=\frac{1}{2}\epsilon_{mks}\frac{\partial u_s}{\partial x_k}=\frac{1}{2}({\rm curl}\,u)_m$, $m=1,2,3$, is the macrorotation vector of the linear theory of elasticity which is kinematically distinct from the microrotation vector.  The internal energy density was chosen in the form
\begin{align}
W&=\frac{1}{2}\Big[(2\mu^*+\kappa)\, \varepsilon_{kl}\varepsilon_{lk}+\lambda\, \varepsilon_{kk}\varepsilon_{ll}+2\,\varkappa\, (r_k-\varphi_k)(r_k-\vartheta_k)+\beta\, \vartheta_{k,l}\vartheta_{l,k}+\gamma\,
\, \vartheta_{l,k}\vartheta_{l,k}+\alpha\, \vartheta_{k,k}\vartheta_{l,l}\Big]
\\
&=\frac{2\mu^*+\kappa}{2}\, \lVert \boldsymbol{\varepsilon}\rVert^2+\frac{\lambda}{2}[\tr(\boldsymbol{\varepsilon})]^2+\varkappa\, \lVert r-\vartheta \rVert^2+\frac{\gamma}{2}\,\lVert\boldsymbol{\mathfrak{K}}\rVert^2+\frac{\beta}{2}\,\bigl\langle \boldsymbol{\mathfrak{K}}, \boldsymbol{\mathfrak{K}}^T\bigr\rangle+\frac{\alpha}{2}\,[\tr(\boldsymbol{\mathfrak{K}})]^2, \notag
\end{align}
where $\boldsymbol{\varepsilon}=\sym\, {\rm D}u$ 
is the classical strain tensor from the linear theory of classical elasticity and $\boldsymbol{\mathfrak{K}} ={\rm D} \vartheta={\rm D} (\axl \mathbf{A}).$

In any case, the notations used by Eringen for the constitutive parameters are coherent and consistent, since we have the alternative form of the stress-strain  relations 
\begin{align}\label{eq:stress_classic1}
{\sigma}_{ji} &=
\mu^*( u_{i,j}+u_{j,i})+\varkappa\, (u_{i,j}-\epsilon_{jis}\vartheta_s)+\lambda \, u_{k,k}\delta_{ji},
\\*
m _{ji}&=\beta\, \vartheta_{j,i}+\gamma\,
\vartheta_{i,j}+ \alpha \,  \vartheta_{k,k}\delta_{ji} 
\, 
\notag
\end{align}
which are agree with the 
 force stress and the couple stress tensor definition   \eqref{eq:stress_classic}. Therefore, we have the same  identifications as those already given.

	The set of these coefficients was used by Lakes in expressing the experimental results in terms of the following  coefficients, called   {\it the micropolar technical constants} by Lakes \cite[Page 2576]{Lakes83}, i.e.,
	\begin{align}
	&\text{Young's modulus } & \qquad  E&=\frac{(2\, \mu^*+\kappa)(3\, \lambda+2\,\mu^*+\kappa)}{2\, \lambda+2\, \mu^*+\kappa}, \notag\\
	&\text{shear modulus } & \qquad  G&=\frac{2\, \mu^*+\kappa}{2},\notag\\
	&\text{Poisson ratio } & \qquad  \nu&=\frac{\lambda}{2\, \lambda+2\, \mu^*+\kappa},\notag\\
	&\text{characteristic length, torsion } & \qquad   \ell_t&=\sqrt{\frac{\beta+\gamma}{2\, \mu^*+\kappa}},\\
	&\text{characteristic length, bending  } & \qquad  \ell_b&=\sqrt{\frac{\gamma}{2\, (2\, \mu^*+\kappa)}},\notag\\
	&\text{coupling number  } & \qquad  N&=\sqrt{\frac{\kappa}{2\, (\mu^*+\kappa)}},\notag\\
	&\text{polar ratio  } & \qquad  \Psi&=\frac{\beta+\gamma}{\alpha+\beta+\gamma}.\notag
	\end{align}

Using the micropolar technical constants, for the materials considered in Lakes' experiments, we will identify the numerical values of the constitutive parameters considered in our formulation (see Table \ref{tab:exper_values_ours}).

\subsection{Mindlin's notation}

Mindlin \cite{mindlin1965stress} has considered the following strain tensors
\begin{align}
\varepsilon_{ij}&=\frac{1}{2}(u_{i,j}+u_{j,i}),\qquad \qquad 
\gamma_{[ij]}=u_{[j,i]}-P_{[ij]},\qquad \qquad 
\kappa_{i[jk]}=P_{[jk],i},
\end{align}
where
\begin{align}
u_{[i,j]}=\frac{1}{2}(u_{i,j}-u_{j,i})=(\skw\, {\rm D} u)_{ij}, \qquad P_{[ij]}=\frac{1}{2}(P_{ij}-P_{ji})=(\skw\, P)_{ij},
\end{align}
i.e., in our notation,
\begin{align}
\boldsymbol{\varepsilon}&=(\varepsilon_{ij})_{i,j=1,2,3}=\sym\, {\rm D} u,\qquad \qquad \boldsymbol{\gamma}=(\gamma_{[ij]})_{i,j=1,2,3}=\skew( {\rm D} u-\mathbf{P}), \\ \boldsymbol{\kappa}&=(\kappa_{i[jk]})_{i,j,k=1,2,3}={\rm D} \skew\, \mathbf{P},\notag
\end{align}
where $\mathbf{P}\in \mathbb{R}^{3\times 3}$ is the microdistortion  tensor arising in the micromorphic theory.
Therefore, using our notation and identifying $\mathbf{A}=\skew\, \mathbf{P}$, we obtain that the strain tensors considered by Mindlin are
actually
\begin{align}
\boldsymbol{\varepsilon}= \sym({\rm D} u-\mathbf{A})=\sym\, \mathbf{e}, \qquad\qquad\qquad \boldsymbol{\gamma}= \skw({\rm D} u-\mathbf{A})=\skw \,\mathbf{e}, \qquad\qquad\qquad \boldsymbol{\kappa}={\rm D} \mathbf{A}.
\end{align}
Note that here $\boldsymbol{\kappa}$ is a third order tensor. 
The isotropic internal energy density in the form considered by Mindlin reads
\begin{align}
W=&\,\mu^{\rm M}\, \varepsilon_{ij}\varepsilon_{ij}+\frac{1}{2}\lambda^{\rm M}\, \varepsilon_{ii}\varepsilon_{jj}+\mu_{\rm c}^{\rm M}\, \gamma_{[ij]}\gamma_{[ij]}+\beta_1^{\rm M}\, \kappa_{i[ik]}\kappa_{j[jk]}+\beta_2^{\rm M}\, \kappa_{i[jk]}\kappa_{i[jk]}+\beta_3^{\rm M}\, \kappa_{i[jk]}\kappa_{j[ik]}\notag\\
=&\,\mu^{\rm M}\, \|\varepsilon\|^2+\frac{\lambda^{\rm M}}{2}\,(\tr (\varepsilon))^2+\mu_{\rm c}^{\rm M}\, \| \gamma\|^2+\beta_1^{\rm M}\, \kappa_{i[ik]}\kappa_{j[jk]}+\beta_2\, \kappa_{i[jk]}\kappa_{i[jk]}+\beta_3^{\rm M}\, \kappa_{i[jk]}\kappa_{j[ik]}\\
=&\,\mu^{\rm M} \,\|\sym({\rm D} u-\mathbf{A})\|^2+\frac{\lambda^{\rm M}}{2}\,(\tr (\sym({\rm D} u-\mathbf{A})))^2+\mu_{\rm c}^{\rm M} \,\| \skw({\rm D} u-\mathbf{A})\|^2\notag\\&\quad +\beta_1^{\rm M}\, \kappa_{i[ik]}\kappa_{j[jk]}+\beta_2^{\rm M}\, \kappa_{i[jk]}\kappa_{i[jk]}+\beta_3^{\rm M}\, \kappa_{i[jk]}\kappa_{j[ik]}.\notag
\end{align}
The assumed isotropic format of the curvature energy is not immediately apparent, since Mindlin deals with a third order tensor. 

The equilibrium equations  in the Mindlin form are
\begin{align}
0&=(\lambda^{\rm M}+\mu^{\rm M}-\mu_{\rm c}^{\rm M})u_{j,ji}+(\mu^{\rm M}+\mu_{\rm c}^{\rm M})u_{i,jj}-2\mu_{\rm c}^{\rm M} P_{[ji],j},\notag\\
0&=(\beta_1^{\rm M}+\beta_3^{\rm M})(P_{[ki],kj}+P_{[jk],ki})+2\beta_2^{\rm M} P_{[ij],kk}-2\mu_{\rm c}^{\rm M} P_{[ij]}+\mu_{\rm c}^{\rm M} (u_{j,i}-u_{i,j}).
\end{align}

We note
\begin{align}
\|\boldsymbol{\kappa}\|^2=\|{\rm D}\mathbf{A}\|^2=2\,\|{\rm D}{\rm axl}(\mathbf{A}) \|^2=2\,\|\boldsymbol{\mathfrak{K}}\|^2&=2\,\|{\rm D} \vartheta\|^2
\end{align}
and that
\begin{align}
\kappa_{i[jk]}=\mathbf{A}_{[jk],i}=\frac{1}{2}\left(A_{jk,i}-A_{kj,i}\right)
=\frac{1}{2}\left(-\epsilon_{jkl}\vartheta_{l,i}+\epsilon_{kjl}\vartheta_{l,i}\right)
=\frac{1}{2}\left(\epsilon_{kjl}+\epsilon_{kjl}\right)\vartheta_{l,i}=\epsilon_{kjl}\,\vartheta_{l,i}.
\end{align}
Therefore, on one hand we deduce
\begin{align}
\kappa_{i[ik]}&=\epsilon_{kil}\,\vartheta_{l,i}=({\rm curl}\, \vartheta)_{k}, \\ \kappa_{i[ik]}\kappa_{j[jk]}&=\|{\rm curl} \,\vartheta\|^2=2 \, \,\|\skw\, {\rm D} \vartheta\|^2=2 \, \|\skw\,  \boldsymbol{\mathfrak{K}}\|^2.\notag
\end{align}
On the other hand we obtain
\begin{align}
\kappa_{i[jk]}\kappa_{i[jk]}&=\epsilon_{kjl}\vartheta_{l,i}\epsilon_{kjm}\vartheta_{m,i}=\epsilon_{kjl}\epsilon_{kjm}\vartheta_{l,i}\vartheta_{m,i}=(\delta_{jj}\delta _{lm}-\delta_{jm}\delta_{jl})\vartheta_{l,i}\vartheta_{m,i}\notag\\
&=(3\,\delta _{lm}-\delta_{lm})\vartheta_{l,i}\vartheta_{m,i}=2\,\delta _{lm}\vartheta_{l,i}\vartheta_{m,i}=2\,\vartheta_{l,i}\vartheta_{l,i}=2\|{\rm D}\vartheta\|^2=2\,\|\boldsymbol{\mathfrak{K}}\|^2\\&=2\left(\|\sym \, \boldsymbol{\mathfrak{K}}\|^2+\|\skw\, \boldsymbol{\mathfrak{K}}\|^2+\frac{1}{3}[\tr (\boldsymbol{\mathfrak{K}})]^2\right)\notag
\end{align}
and
\begin{align}
\kappa_{i[jk]}\kappa_{j[ik]}&=\epsilon_{kjl}\vartheta_{l,i}\epsilon_{kim}\vartheta_{m,j}=\epsilon_{kjl}\,\epsilon_{kim}\,\vartheta_{l,j}\vartheta_{m,i}\notag\\
&=(\delta_{ji}\delta _{lm}-\delta_{jm}\delta_{li})\vartheta_{l,i}\vartheta_{m,j}=\,\vartheta_{m,i}\vartheta_{m,i}-\vartheta_{i,i}\vartheta_{j,j}=\|{\rm D}\vartheta\|^2-[\tr ({\rm D}\vartheta)]^2\notag\\
&=\|\sym \, \boldsymbol{\mathfrak{K}}\|^2+\|\skw \, \boldsymbol{\mathfrak{K}}\|^2+\frac{1}{3}[\tr (\boldsymbol{\mathfrak{K}})]^2-[\tr ({\rm D}\vartheta)]^2\\
&=\|\sym\, \boldsymbol{\mathfrak{K}}\|^2+\|\skw \, \boldsymbol{\mathfrak{K}}\|^2-\frac{2}{3}[\tr (\boldsymbol{\mathfrak{K}})]^2.\notag
\end{align}

Thus, the internal energy density  considered by Mindlin can be rewritten in the form 
\begin{align}
W
=&\,\mu^{\rm M}\,\|\sym({\rm D} u-\mathbf{A})\|^2+\frac{\lambda^{\rm M}}{2}\,(\tr (\sym({\rm D} u-\mathbf{A})))^2+\mu_{\rm c} \,\| \skw({\rm D} u-\mathbf{A})\|^2\notag\\&\quad  +2\,\beta_1^{\rm M}\, \|\skw\,  \boldsymbol{\mathfrak{K}}\|^2+2\,\beta_2^{\rm M}\left(\|\sym \boldsymbol{\mathfrak{K}}\|^2+\|\skw \boldsymbol{\mathfrak{K}}\|^2+\frac{1}{3}[\tr (\boldsymbol{\mathfrak{K}})]^2\right)\notag\\&\quad +\beta_3^{\rm M}\, \left(\|\sym \boldsymbol{\mathfrak{K}}\|^2+\|\skw \boldsymbol{\mathfrak{K}}\|^2-\frac{2}{3}[\tr (\boldsymbol{\mathfrak{K}})]^2\right)\notag\\
=&\,\mu^{\rm M}\,\lVert\dev   \, \sym\,\mathbf{e}\rVert ^{2}+\mu_{\rm c}^{\rm M}\,\lVert \skw\,\mathbf{e}\rVert ^{2}+\frac{2\mu^{\rm M} +3\,\lambda^{\rm M} }{6}\left[\mathrm{tr} \left(\mathbf{e}\right)\right]^{2}\\&\quad  +(2\,\beta_2^{\rm M}+\beta_3^{\rm M})\|\sym \, \boldsymbol{\mathfrak{K}}\|^2 +(2\,\beta_1^{\rm M}+2\,\beta_2^{\rm M}+\beta_3^{\rm M})\, \|\skw\,  \boldsymbol{\mathfrak{K}}\|^2+ \frac{2\,\beta_2^{\rm M}-2\,\beta_3^{\rm M}}{3}[\tr (\boldsymbol{\mathfrak{K}})]^2\notag\\
=&\,\mu^{\rm M}\,\lVert\dev   \, \sym\,\mathbf{e}\rVert ^{2}+\mu_{\rm c}^{\rm M}\,\lVert \skw\,\mathbf{e}\rVert ^{2}+\frac{2\mu^{\rm M} +3\,\lambda^{\rm M} }{6}\left[\mathrm{tr} \left(\mathbf{e}\right)\right]^{2}\notag\\& \quad +(2\,\beta_2^{\rm M}+\beta_3^{\rm M})\|\dev \,\sym \, \boldsymbol{\mathfrak{K}}\|^2 +(2\,\beta_1^{\rm M}+2\,\beta_2^{\rm M}+\beta_3^{\rm M})\, \|\skw\,  \boldsymbol{\mathfrak{K}}\|^2+ \frac{4\,\beta_2^{\rm M}-\,\beta_3^{\rm M}}{3}[\tr (\boldsymbol{\mathfrak{K}})]^2\notag.
\notag
\end{align}

By comparing the Mindlin's internal energy density to the form \eqref{enersplit} considered by us in the new formulation, we obtain the identification of the parameters
\begin{align}
\lambda_{\text{e}} &=\lambda^{\rm M} \, ,\qquad \qquad \qquad \qquad \ \ \, 
\,\,\mu_{\text{e}} =\mu^{\rm M}\, ,\qquad \qquad \qquad \qquad \qquad \qquad 
\,\mu_{\text{c}} =\mu_{\text{c}} ^{\rm M}, \notag\\
\alpha_1 &=\frac{ 2}{L_{\rm c}^2 \, \mu^{\rm M}} (2\,\beta_2^{\rm M}+\beta_3^{\rm M})\, ,\qquad 
\alpha_2=\frac{ 2}{L_{\rm c}^2 \, \mu^{\rm M}} (2\,\beta_1^{\rm M}+2\,\beta_2^{\rm M}+\beta_3^{\rm M}) ,\qquad 
\alpha_3 =\frac{ 4}{3\,L_{\rm c}^2 \, \mu^{\rm M}} (2\,\beta_2^{\rm M}-3\,\beta_3^{\rm M}) \, 
\end{align}
and, vice versa,
\begin{align}
\lambda^{\rm M}&=\lambda_{\text{e}}  \, ,\qquad \qquad \qquad \ \ \ \,\, \,
\,\,\mu^{\rm M}=\mu_{\text{e}} \, ,\qquad \qquad \quad \qquad \quad  \,
\,\,\,\,\mu_{\text{c}} ^{\rm M}=\mu_{\text{c}} , \notag\\
\beta_1^{\rm M}&=\frac{L_{\rm c}^2 \, \mu_{\text{e}}}{4}(  \alpha_2 -\alpha_1)\, ,\qquad 
\beta_2^{\rm M}=\frac{3\,L_{\rm c}^2 \, \mu_{\text{e}}}{4}( 2\, \alpha_1 +\alpha_3),\qquad 
\beta_3^{\rm M}=\frac{L_{\rm c}^2 \, \mu_{\text{e}}}{8}(  \alpha_1 -\frac{3}{2}\alpha_3)\, .
\end{align}

\subsection{Nowacki's notation}
In \cite{hassanpour2017micropolar}  the Nowacki notation \cite{nowacki1972theory} is used and  another set of parameters is presented.  The constitutive relations are given in  indices as\footnote{The tensors ${\sigma}_{ji}$ and $m _{ji}^*$ must be \textit{transposed} before the parameters' identification with the ones proposed in this paper.}
\begin{align}\label{eq:stress_classic}
{\sigma}_{ji} &=
(\mu^{\rm N}+\varkappa^{\rm N}) \, ( u_{i,j}-  \epsilon_{jim}\vartheta_m)+ (\mu^{\rm N}-\varkappa^{\rm N})\,(u_{j,i}-  \epsilon_{ijm}\vartheta_m)+\lambda ^{\rm N} \,  u_{k,k}\delta_{ji},
\\
m _{ji}^*&=
(\gamma^{\rm N}+\beta^{\rm N}) \, \vartheta_{i,j}+(\gamma^{\rm N}-\beta^{\rm N}) \, \vartheta_{j,i}+ \alpha^{\rm N}  \, \vartheta_{k,k}\delta_{ji} 
\, ,
\notag
\end{align}
This implies the following identifications
\begin{align}
\mu^*+\varkappa&=\mu^{\rm N}+\varkappa^{\rm N},\qquad \mu^*=\mu^{\rm N}-\varkappa^{\rm N},\qquad \lambda =\lambda^{\rm N}, \\
\beta &=\gamma^{\rm N}-\beta^{\rm N}, \qquad \ \ \gamma =\gamma^{\rm N}+\beta^{\rm N}, \qquad\alpha =\alpha^{\rm N} \,,\notag
\end{align}
and
\begin{align}
\mu^{\rm N}&=\mu^*+\frac{\varkappa}{2},\qquad \varkappa^{\rm N}=\frac{\varkappa}{2},\qquad \qquad\,\lambda^{\rm N}=\lambda , \\
\gamma^{\rm N}&=\frac{\beta + \gamma }{2},\qquad\ \ \beta^{\rm N}=\frac{ \gamma-\beta }{2}, \ \ \ \ \ \ \  \alpha^{\rm N}=\alpha  \,.\notag
\end{align}
Vice versa, it holds
\begin{align}
\mu^*&=\mu^{\rm N}-\varkappa^{\rm N},\qquad\, \varkappa=2\varkappa^{\rm N},\qquad\qquad \lambda=\lambda^{\rm N}, \\
\beta &=\gamma^{\rm N}-\beta^{\rm N}, \qquad \ \gamma =\gamma^{\rm N}+\beta^{\rm N}, \,\qquad\alpha =\alpha ^{\rm N}\,.\notag
\end{align}

By using the identifications given in Section \ref{idsec}, we obtain the comparison between the constitutive parameters used by Nowacki and the parameters used in the dislocation format of the internal energy density, i.e.
\begin{align}
\lambda_{\text{e}} &=\lambda^{\rm N} \, ,\qquad \qquad \quad \ \,
\,\,\mu_{\text{e}} =\mu^{\rm N}\, ,\qquad \qquad \quad
\!\mu_{\text{c}} =\varkappa^{\rm N}\, ,\notag\\
\alpha_1 &=\frac{ 2}{L_{\rm c}^2 \, \mu^{\rm N}}  \gamma^{\rm N}\, ,\qquad \quad 
\alpha_2=\frac{ 2}{L_{\rm c}^2 \, \mu^{\rm N}}  \beta^{\rm N} ,\qquad\, 
\alpha_3 =\frac{ 2}{L_{\rm c}^2 \, \mu^{\rm N}}  \alpha^{\rm N} \, .
\end{align}

\section{Linear Cosserat elasticity as a particular case of the  relaxed \\ micromorphic model}\label{rel}\setcounter{equation}{0}
The form of the internal energy proposed here keeps the form of the internal energy proposed in a more general theory, namely, the relaxed micromorphic theory. In  the micromorphic theory, the microdistortion tensor $ \mathbf{P}=( \mathbf{P}_{ij}):\Omega\times [0,T]\rightarrow \mathbb{R}^{3 \times 3}$  describes the substructure of the material which can rotate, stretch, shear and shrink, while $u=(u_i) :\Omega\times [0,T]\rightarrow  \mathbb{R}^3$  is the displacement of the macroscopic material points. In the relaxed micromorphic model, in which the  appearance of a strictly positive Cosserat modulus $\mu_{\rm c} >0$ is related to the isotropic Eringen-Claus  model for dislocation dynamics \cite{Eringen_Claus69,EringenClaus,Eringen_Claus71}, {the free energy is given by }
\begin{align}\label{XXXX}
W_{\rm relax}&=\mu_{\rm e}  \|\sym \,(\mathrm{D}u -\mathbf{P})\|^2+\mu_{\rm c} \|\skw(\mathrm{D}u -\mathbf{P})\|^2+ \frac{\lambda_{\rm e} }{2}\, [\tr(\mathrm{D}u -\mathbf{P})]^2+\mu_{\rm micro} \|\sym \, \mathbf{P}\|^2+ \frac{\lambda_{\rm micro}}{2} [\tr(\mathbf{P})]^2\notag\\&
\quad \quad  +\frac{\mu L_{\rm c}^2}{2}\left[{a_1}\|\dev  \,\sym \,\Curl\, \mathbf{P}\|^2 +{a_2}\| \skw \,\Curl\, \mathbf{P}\|+ \frac{a_3}{3}\, \tr(\Curl\, \mathbf{P})^2\right],
\end{align}
where $(\mu_{\rm e} ,\lambda_{\rm e} ) $,  $(\mu_{\rm micro},\lambda_{\rm micro})$,  $\mu_{\rm c}, L_{\rm c} $  and $( a_1, a_2, a_3)$  are the elastic moduli  representing the parameters related to the meso-scale,  the
parameters related to the micro-scale, the Cosserat couple modulus $\mu_{\rm c}\geq 0$, the characteristic length $L_{\rm c}$, and the three
general isotropic curvature parameters (non dimensional weights), respectively.  

Due to the generalized Korn-type inequalities \cite{agn_neff2015poincare,agn_neff2011canonical,agn_neff2012canonical,agn_neff2012maxwell,agn_bauer2014new,agn_bauer2013dev}, the relaxed micromorphic model is well-posed for either
\begin{align}\label{ew}
\mu_{\rm e}&>0, \qquad \qquad \quad\, 3\,\lambda_{\rm e}+2\,\mu_{\rm e}>0, \qquad\qquad \mu_{\rm c}\geq 0,\notag\\
\mu_{\rm micro}&>0, \qquad 3\,\lambda_{\rm micro}+2\,\mu_{\rm micro}>0, \\  a_1&>0, \qquad \qquad\qquad \qquad \ \ a_2\geq 0, \qquad \qquad a_3\geq 0\notag
\end{align}
or
\begin{align}
\mu_{\rm e}&>0, \qquad \qquad \quad\, 3\,\lambda_{\rm e}+2\,\mu_{\rm e}>0, \qquad\qquad \mu_{\rm c}\geq 0,\notag\\
\mu_{\rm micro}&>0, \qquad 3\,\lambda_{\rm micro}+2\,\mu_{\rm micro}=0, \\  a_1&>0, \qquad \qquad\qquad \qquad \ \ a_2>0, \qquad \qquad a_3> 0.\notag
\end{align}

Note that here, setting the Cosserat couple modulus $\mu_{\rm c}=0$ is possible since the two independent fields $u$ and $\mathbf{P}$  are still constitutively coupled via their strain tensors ${\rm sym}\,{\rm D}u$ and ${\rm sym}\,P$ (this strain tensor coupling is missing in the Cosserat framework). The linear Cosserat model appears  as a singular limit  of the relaxed micromorphic model for
\begin{align}\label{limit}
\mu_{\rm micro}, \qquad 3\lambda_{\rm micro}+2\mu_{\rm micro}\to \infty.
\end{align}
In this sense one can say that the relaxed micromorphic model uses a straightforward  extension of the curvature energy in the Cosserat model (once the Cosserat model is represented in the appropriate second order dislocation tensor format). The Cosserat model is often supposed to describe a ``rigid microstructure". This interpretation perfectly fits with the limit \eqref{limit} constraining effectively the affine microdistortion $P$ to infinitesimal rigid motions $\mathbf{P}=\mathbf{A}\in \so(3)$ and giving back the Cosserat model.

\subsection{Identification of the relaxed micromorphic parameters in Mindlin's notation}

As a final identification, we report the correspondence between the coefficients in the relaxed micromorphic  model and in the classical Mindlin's notation for linear elasticity with microstructure \cite{Mindlin62}. The isotropic Mindlin formulation features $7+15=22$ number of coefficients, while the isotropic relaxed micromorphic model displays altogether $5+3=8$ coefficients. The relaxed micromorphic model is nevertheless a subclass of Mindlin's formulation. 
The identification for the coefficients of the linear part of the energy is (see  \cite{neff2004material,mindlin1963microstructure})

\begin{equation}
\begin{array}{rlrlrlrlrlrl}
\mu^{\rm M} &= \mu_{\rm micro} \, ,
&
\qquad
\lambda^{\rm M} &= \lambda_{\rm micro} \, ,
&
\qquad
g_1^{\rm M} &= - \lambda_{\rm micro} \, ,
\qquad
g_2^{\rm M} = -2\mu_{\rm micro} \, ,
\\
b_{1}^{\rm M} &= \lambda_{\rm e} + \lambda_{\rm micro} \, ,
&
\qquad
b_{2}^{\rm M} &= \mu_{\rm e} +  \mu_{\rm micro} +  \mu_{\rm c} \, ,
&
\qquad
b_{3}^{\rm M} &= \mu_{\rm e} +  \mu_{\rm micro} -  \mu_{\rm c} \, ,
\notag
\end{array}
\end{equation}

while for the curvature parameters we have (see \cite{rizzi2021bending,mindlin1963microstructure})
\begin{align}
	a^{\rm M}_{1,2,3,5,8,11,14,15} &= 0 \, ,
	\qquad a^{\rm M}_{4} = \mu \, L_c^2 \frac{2a_3-a_1}{3} \, ,
	\qquad a^{\rm M}_{10} = \mu \, L_c^2 \frac{a_1 + a_2}{2} \, ,
	\qquad a^{\rm M}_{13} = \mu \, L_c^2 \frac{a_1 - a_2}{2}
	\, .
\end{align}

\section{Nonlinear isotropic Cosserat model}\label{Cos}\setcounter{equation}{0}

The nonlinear deformation of the body occupying the domain $\Omega$ is described by a map $\varphi$ (\textit{called deformation}) and by a \textit{microrotation} orthogonal tensor field $\overline{\mathbf{R}}$,
\begin{equation}
\varphi:\Omega\subset\mathbb{R}^3\rightarrow\mathbb{R}^3, \qquad  \ \overline{\mathbf{R}}:\Omega\subset\mathbb{R}^3\rightarrow {\rm SO}(3)\, .
\end{equation}
We denote the current configuration by $\Omega:=\varphi(\Omega)\subset\mathbb{R}^3$. 
The deformation and the microrotation is solution of the following \textit{geometrically nonlinear minimization problem} posed on $\Omega$:
\begin{equation}\label{minprob}
I(\varphi,\overline{\mathbf{R}})=\dd\int_{\Omega}\left[W_{\rm{mp}}(\overline {\mathbf{U}} )+
W_{\rm{curv}}(\boldsymbol{\alpha})\right]dV
{\to}
\textrm{\ \ min.} \quad  {\rm   w.r.t. }\quad (\varphi,\overline{\mathbf{R}})\, ,
\end{equation}
where
\begin{align}
\mathbf{F}:\,=\,&{\rm D}\varphi\in\mathbb{R}^{3\times3}\, \qquad \qquad \qquad \qquad \qquad \textrm{(the deformation gradient)},  \notag\\
\overline {\mathbf{U}} :\,=\,&\dd\overline{\mathbf{R}}^T \mathbf{F}\in\mathbb{R}^{3\times3} \ \ \  \ \, \, \  \qquad \qquad \qquad \quad \ \textrm{(the non-symmetric Biot-type stretch tensor)},  \notag\\
\boldsymbol{\alpha}:\,=\,&\overline{\mathbf{R}}^T\, \Curl \,\overline{\mathbf{R}}\in\mathbb{R}^{3\times 3} \,  \qquad \qquad\qquad  \quad  \textrm{(the second order  dislocation density tensor)}\, ,  \\
\dd W_{\rm{mp}}(\overline{ \mathbf{U}}):\,=\,&\dd\mu_{\rm e}\,\lVert \text{sym}(\overline {\mathbf{U}} -\id_3)\rVert^2+\mu_{\rm c}\,\lVert \text{skew}(\overline {\mathbf{U}} -\id_3)\rVert^2+
\dd\frac{\lambda_{\rm e}}{2}\,[{\rm tr}(\text{sym}(\overline {\mathbf{U}} -\id_3))]^2\ \ \, \textrm{(physically linear)}\, ,  \notag\\
\dd W_{\rm{curv}}( \boldsymbol{\alpha}):\,=\,&\mu_{\rm e}\,\frac{{L}_{\rm c}^2}{2}\left( a_1\,\lVert\dev  \,\text{sym}\, \boldsymbol{\alpha}\rVert^2+a_2\,\lVert \text{skew}\, \boldsymbol{\alpha}\rVert^2+  \frac{a_3}{3}\,
[{\rm tr}(\boldsymbol{\alpha})]^2\right)\qquad \qquad\qquad \textrm{(curvature energy)}, \notag
\end{align}
and $dV$ denotes the  volume element in the $\Omega$-configuration. 

The total elastically stored energy $W=W_{\rm mp}+W_{\rm curv}$ depends on
the deformation gradient $\mathbf{F}={\rm D}\varphi$ and microrotations $\overline{ \mathbf{R}}$ together with their spatial
derivatives. In general, the \textit{{Biot-type stretch tensor}} $\overline{ \mathbf{U}}$ is not symmetric (the first Cosserat deformation tensor \cite{Cosserat09}). The parameters $\mu_{\rm e}$ and $\lambda_{\rm e}$ are the \textit{Lam\'e constants}
of classical isotropic elasticity, $\frac{2\mu_{\rm e}+3\lambda_{\rm e}}{3}$ is the \textit{infinitesimal bulk modulus}, $a_1, a_2, a_3$ are \textit{non-dimensional constitutive curvature coefficients (weights)}, $\mu_{\rm c}\geq 0$ is called the \textit{{Cosserat couple modulus}} and ${L}_{\rm c}>0$ introduces an \textit{{internal length}} which is {characteristic} for the material, e.g., related to the grain size in a polycrystal. The
internal length ${L}_{\rm c}>0$ is responsible for \textit{size effects} in the sense that smaller samples are relatively stiffer than
larger samples. 

The nonlinear Cosserat model is known to be well-posed \cite{neff2015existence,Neff_micromorphic_rse_05,Neff_Gamm04,LNPzamp2013} for
\begin{align}
\mu_{\rm e}>0, \qquad 2\mu_{\rm e}+3\lambda_{\rm e}>0, \qquad \mu_{\rm c}>0, \qquad a_1>0, \qquad a_2>0, \qquad a_3> 0. 
\end{align}

For the rotation tensor $ \overline{\mathbf{R}}\in\rm{SO}(3) $ there exists a unique  skew-symmetric matrix  \begin{align}
{\mathbf{A}}=\mathbf{Anti}(\vartheta_1,\vartheta_2,\vartheta_3):=\begin{footnotesize}
\begin{pmatrix}
0&-\vartheta_3&\vartheta_2\\
\vartheta_3&0&-\vartheta_1\\
-\vartheta_2&\vartheta_1&0
\end{pmatrix}\end{footnotesize}\in \mathfrak{so}(3),
\end{align}
 such that 
\begin{align}\label{lxQ} \overline{\mathbf{R}}=\exp\,\mathbf{A}\;= \;\sum_{k=0}^{\infty} \frac{1}{k!} \,\mathbf{A}^k\; = \;\id_3 + \mathbf{A}+\textrm{h.o.t.}\end{align}

Considering the linearisation, i.e. for situations of small  deformations and small curvature, 
\begin{align}
\varphi={\rm id}+u+\textrm{h.o.t.}, \qquad  \overline{\mathbf{R}}=\exp
\,\mathbf{A}
\, = \;\id_3 + \mathbf{A}+\textrm{h.o.t.}
\end{align}
with $u:\Omega\to \mathbb{R}^3$ the infinitesimal displacement and the tensor field $\mathbf{A}$   the infinitesimal microrotation, the linear Cosserat model is obtained. 
Here, ``h.o.t" stands for terms of order higher than linear with respect to $u$ and $\mathbf{A}$.

Note that, due to the results obtained in \cite{Neff_curl06}, the curvature tensor $\boldsymbol{\alpha}=\overline{\mathbf{R}}^T\, \Curl \,\overline{\mathbf{R}}\in\mathbb{R}^{3\times 3}$ controls all partial derivatives\footnote{For example, we can express the isotropic term \begin{align}
\|\overline{\mathbf{R}}^T{\rm D}\overline{\mathbf{R}}\|^2=\|{\rm D}\overline{\mathbf{R}}\|^2=	2\,\left( \,\|\dev\,\text{sym} \, \boldsymbol{\alpha} \|^2+\,\|\text{skew} \, \boldsymbol{\alpha} \|^2+\frac{1}{12}\,
	[{\rm tr}(\boldsymbol{\alpha} )]^2\right).
	\end{align}} of $\overline{\mathbf{R}}$. It is important to realize that,  contrary to the linear Cosserat model, the nonlinear Cosserat model allows to set the Cosserat couple modulus $\mu_{\rm c}=0$. In this case the curvature energy needs to be modified to allow a control of 
$\boldsymbol{\alpha}=\overline{\mathbf{R}}^T\, \Curl \,\overline{\mathbf{R}}\in\mathbb{R}^{3\times 3}$ in ${\rm L}^p(\Omega)$, $p>n$, i.e.
\begin{align}
  W_{\rm{curv}}( \boldsymbol{\alpha}):\,=\,&\mu_{\rm e}\,\left[\frac{{L}_{\rm c}^2}{2}\left( a_1\,\lVert\dev  \,\text{sym}\, \boldsymbol{\alpha}\rVert^2+a_2\,\lVert \text{skew}\, \boldsymbol{\alpha}\rVert^2+  \frac{a_3}{3}\,
 [{\rm tr}(\boldsymbol{\alpha})]^2\right)\right]^{\frac{p}{2}}.
 \end{align}
 
 \begin{table}[h!]\begin{center}
 		\resizebox{17cm}{!}{	\begin{tabular}{ |c|c| c | c |c|c|}\hline 
 				Name
 				&\begin{minipage}{1.5cm}\medskip classical elasticity \medskip \end{minipage}
 				&Our notations
 				& \begin{minipage}{2cm}\medskip Lakes  \\(Eringen)\medskip \end{minipage}
 				&Mindlin &\begin{minipage}{2cm}\medskip Hassanpour \&\,Heppler (Nowacki)\medskip \end{minipage}
 				\\\hline 
 				\begin{minipage}{3cm}\medskip $G$ (shear modulus)\medskip \end{minipage}
 				&  $\mu$
 				&  $\mu_{\rm e}$ 
 				&  $\mu^* +\dfrac{\varkappa}{2}$
 				&  $\mu^{\rm M}$
 				&  $\mu^{\rm N}$
 				\\\hline 
 				\begin{minipage}{3cm}\medskip Lam\'{e} \\ first parameter \medskip \end{minipage}
 				&  $\lambda$
 				&  $\lambda_{\rm e}$ 
 				&  $\lambda$
 				&  $\lambda^{\rm M}$
 				&  $\lambda^{\rm N}$ 
 				\\\hline 
 				\begin{minipage}{3cm}\medskip 	$\nu$ (classical \\ Poisson ratio)\medskip\end{minipage}
 				&  $\nu$
 				&  $\dfrac{\lambda_{\rm e}}{2(\mu_{\rm e}+\lambda_{\rm e})}$
 				&  $\dfrac{\lambda}{2(\lambda +\mu^*+\dfrac{\varkappa}{2})}$
 				&  $\dfrac{\lambda^{\rm M}}{2(\mu^{\rm M}+\lambda^{\rm M})}$
 				&  $\dfrac{\lambda^{\rm N}}{2(\mu^{\rm N}+\lambda^{\rm N})}$
 				\\\hline 
 				\begin{minipage}{3cm}\medskip $E$ (classical \\ Young's modulus)\medskip\end{minipage}
 				&  $E$
 				&  $\dd\frac{\mu_{\rm e}(2\,\mu_{\rm e}+3\, \lambda_{\rm e})}{\mu_{\rm e}+\lambda_{\rm e}}$
 				&  $\dd\frac{(2\, \mu^*+\varkappa)(3\, \lambda+2\,\mu^*+\varkappa)}{2\, \lambda+2\, \mu^*+\varkappa}$
 				&  $\dd\frac{\mu^{\rm M}(2\,\mu^{\rm M}+3\, \lambda^{\rm M})}{\mu^{\rm M}+\lambda^{\rm M}}$
 				&  $\dd\frac{\mu^{\rm N}(2\,\mu^{\rm N}+3\, \lambda^{\rm N})}{\mu^{\rm N}+\lambda^{\rm N}}$
 				\\\hline 
 				\begin{minipage}{3cm}\medskip $\kappa$ (classical \\ bulk modulus)\medskip\end{minipage}
 				&  $\kappa$
 				&  $\lambda_{\rm e}+\dfrac{2}{3}\mu_{\rm e}$
 				&  $\lambda + \dfrac{2}{3}\left(\mu^* + \dfrac{\varkappa}{2}\right)$
 				&  $\lambda^{\rm M}+\dfrac{2}{3}\mu^{\rm M}$
 				&  $\lambda^{\rm N}+\dfrac{2}{3}\mu^{\rm N}$
 				\\\hline 
 				\begin{minipage}{3cm}\medskip mass density\medskip \end{minipage}
 				&  $\rho$
 				&  $\rho$ 
 				&  $\rho$ 
 				&  $\rho$ 
 				&  $\rho$
 				\\\hline 
 				\begin{minipage}{3cm}\medskip Cosserat \\ couple modulus\medskip \end{minipage}
 				&  --
 				&  $\mu_{\rm c}$ 
 				&  $\dfrac{\varkappa}{2}$  
 				&  $\mu_{\rm c}^{\rm M}$ 
 				&  $\varkappa^{\rm N}$ 
 				\\\hline 
 				\begin{minipage}{3cm}\medskip first Cosserat \\ twist coefficient\medskip \end{minipage}
 				&  --
 				&  $\alpha_1 = a_1$  
 				&  $\dfrac{\gamma +\beta }{L_{\rm c}^2 \left(\mu ^*+\dfrac{\varkappa}{2}\right)}$
 				&  $\dfrac{2}{\mu^{\rm M} \, L_{\rm c}^2}(2 \beta_{2}^{\rm M}+\beta_{3}^{\rm M})$
 				&  $\dfrac{2\gamma^{\rm N}}{\mu^{\rm N}\,L_{\rm c}^2}$
 				\\\hline 
 				\begin{minipage}{3cm}\medskip second Cosserat \\ twist coefficient\medskip \end{minipage}
 				&  --
 				&  $\alpha_2 = a_2$  
 				&  $\dfrac{\gamma -\beta }{L_{\rm c}^2 \left(\mu ^*+\dfrac{\varkappa}{2}\right)}$ 
 				&  $\dfrac{2}{\mu^{\rm M} \, L_{\rm c}^2}(2 (\beta_{1}^{\rm M}+\beta_{2}^{\rm M})+\beta_{3}^{\rm M})$
 				&  $\dfrac{2\beta^{\rm N}}{\mu^{\rm N}\,L_{\rm c}^2}$
 				\\\hline 
 				\begin{minipage}{3cm}\medskip third Cosserat \\ twist coefficient\medskip \end{minipage}
 				&  --
 				&  $\alpha_3 = \dfrac{2}{3}\left(4a_3-a_1\right)$  
 				&  $\dfrac{2 \alpha }{L_{\rm c}^2 \left(\mu ^*+\dfrac{\varkappa}{2}\right)}$ 
 				&  $\dfrac{2}{\mu^{\rm M} \, L_{\rm c}^2}\dfrac{4(\beta_{2}^{\rm M}-\beta_{3}^{\rm M})}{3}$
 				&  $\dfrac{2\alpha^{\rm N}}{\mu^{\rm N}\,L_{\rm c}^2}$
 				\\\hline 
 				\begin{minipage}{3cm}\medskip $\Psi$ (dimensionless \\ polar ratio) \medskip \end{minipage}
 				&  --
 				&  $\dd\frac{2\,\alpha_1}{\alpha_3+2\,\alpha_1}=\frac{3 \,a_1}{2\, a_1+4\, a_3}$
 				&  $\dd\frac{\beta+\gamma}{\alpha+\beta+\gamma}$
 				&  $\dfrac{3 (2 \beta_{2}^{\rm M}+\beta_{3}^{\rm M})}{10 \beta_{2}^{\rm M}-\beta_{3}^{\rm M}}$
 				&  $\dd\frac{2\,\gamma^{\rm N}}{\alpha^{\rm N}+2\,\gamma^{\rm N}}$
 				\\\hline 
 				\begin{minipage}{3cm}\medskip $\ell_t$ (characteristic \\ length for torsion) \medskip\end{minipage}
 				&  --
 				&  $L_{\rm c}\sqrt{\dfrac{\alpha_1}{2}}=L_{\rm c}\sqrt{\dfrac{a_1}{2}}$
 				&  $\dd\sqrt{\frac{\beta+\gamma}{2\, \mu^*+\varkappa}}$
 				&  $\sqrt{\dfrac{2 \beta_{2}^{\rm M}+\beta_{3}^{\rm M}}{\mu^{\rm M}}}$
 				&  $\dd\sqrt{\frac{\gamma^{\rm N}}{\mu^{\rm N}}}$
 				\\\hline 
 				\begin{minipage}{3cm}\medskip $\ell_b$ (characteristic \\ length for bending) \medskip\end{minipage}
 				&  --
 				&  $L_{\rm c}\sqrt{\dfrac{\alpha_1+\alpha_2}{8}}$=$L_{\rm c}\sqrt{\dfrac{a_1+a_2}{8}}$
 				&  $\dd\sqrt{\frac{\gamma}{2\, (2\, \mu^*+\varkappa)}}$
 				&  $\sqrt{\dfrac{\beta_{1}^{\rm M}+2 \beta_{2}^{\rm M}+\beta_{3}^{\rm M}}{2 \mu^{\rm M}}}$
 				&  $\dd\sqrt{\frac{\beta^{\rm N}+\gamma^{\rm N}}{4\,\mu^{\rm N}}}$
 				\\\hline 
 				\begin{minipage}{3cm}\medskip $N$ (Cosserat \\ coupling number) \medskip\end{minipage}
 				&  --
 				&  $\dd\sqrt{\frac{\mu_{\rm c}}{\mu_{\rm e}+\mu_{\rm c}}}$
 				&  $\dd\sqrt{\frac{\varkappa}{2\, (\mu^*+\varkappa)}}$
 				&  $\dd\sqrt{\frac{\mu_{\rm c}^{\rm M}}{\mu^{\rm M}+\mu_{\rm c}^{\rm M}}}$
 				&  $\dd\sqrt{\frac{\varkappa^{\rm N}}{\mu^{\rm N}+\varkappa^{\rm N}}}$
 				\\\hline 
 				\begin{minipage}{3cm}\medskip $\xi$ (micropolar twist \\ Poisson's ration) \medskip\end{minipage}
 				&  --
 				&  $\dfrac{\alpha_3}{2(\alpha_1+\alpha_3)}$
 				&  $\dfrac{\alpha }{2\, \alpha +\beta +\gamma }$
 				&  $\dfrac{2 (\beta_{3}^{\rm M}-\beta_{2}^{\rm M})}{\beta_{3}^{\rm M}-10 \beta_{2}^{\rm M}}$
 				&  $\dfrac{\alpha^{\rm N}}{2(\gamma^{\rm N}+\alpha^{\rm N})}$
 				\\\hline 
 				\begin{minipage}{3cm}\medskip $\mathcal{E}$ (micropolar \\ tortile or torsional \\ modulus)\medskip\end{minipage}
 				&  --
 				&  $\dfrac{\mu_{\rm e}\, L_{\rm c}^2}{2}\dfrac{\alpha_1(2\alpha_1+3\alpha_3)}{\alpha_1+\alpha_3}$
 				&  $\dfrac{(\beta +\gamma ) (3\, \alpha +\beta +\gamma )}{2 \alpha +\beta +\gamma }$
 				&  $6 \left(\dfrac{\beta_{3}^{\rm M^2}-8 \beta_{2}^{\rm M^2}-2 \beta_{2}^{\rm M} \beta_{3}^{\rm M}}{\beta_{3}^{\rm M}-10 \beta_{2}^{\rm M}}\right)$
 				&  $\dfrac{\gamma^{\rm N}(2\gamma^{\rm N}+3\alpha^{\rm N})}{\gamma^{\rm N}+\alpha^{\rm N}}$
 				\\\hline 
 				\begin{minipage}{3cm}\medskip $\mathcal{B}$ (micropolar \\ tortile or torsional \\ bulk modulus)\medskip\end{minipage}
 				&  --
 				&  $\dfrac{\mu_{\rm e}\, L_{\rm c}^2}{2}\dfrac{2\alpha_1+3\alpha_3}{3}$
 				&  $\dfrac{(3 \alpha +\beta +\gamma )}{3}$
 				&  $\dfrac{2}{3} (4 \beta_{2}^{\rm M}-\beta_{3}^{\rm M})$
 				&  $\alpha^{\rm N}+\dfrac{2}{3}\gamma^{\rm N}$
 				\\\hline 
 				\begin{minipage}{3cm}\medskip microinertia density \medskip \end{minipage}
 				&  --
 				&  $\eta$ 
 				&  $\dfrac{j}{\tau_{\rm c}^2 (\mu^* +\dfrac{\varkappa}{2})}$ 
 				&  $\dfrac{j}{\tau_{\rm c}^2 \mu^{\rm M} }$ 
 				&  $\dfrac{j}{\tau_{\rm c}^2 \mu^{\rm N} }$ 
 				\\\hline 
 		\end{tabular}}
 	\end{center}
 	\caption{A concluding table for converting all the needed quantities in terms of the new constitutive parameters. The terminology is that used in \cite{hassanpour2017micropolar}. Note that $\Psi=\frac{3}{2}$ if and only if $a_3=0$.}
 	\label{tabelC}
 \end{table}

 We also note that for $\mu_{\rm c}>0$ the quadratic nonlinear Cosserat model is {\bf redundant}  in the sense of Romano et al.  \cite{romano2016micromorphic}. Redundancy is connected to the way rigid body movements are penalised in the elastic energy. Certainly, zero elastic energy occurs if and only if the body undergoes a rigid body movement. In the case of the Cosserat model, however, a part of the elastic energy being zero already suffices to constrain the movement to a rigid motion. In this sense, \textbf{the typical Cosserat model is over-constraining rigid movements}. More precisely, for positive Cosserat couple modulus  $\mu_{\rm c}>0$ it holds
 \begin{align}
  W_{\rm{mp}}( \overline{ \mathbf{U}})=0\quad \Rightarrow\quad W_{\rm{curv}}( \boldsymbol{\alpha})=0
 \end{align}
 and the strain measure already determines the curvature measure in zero.
 To see this, observe that for $\mu_{\rm c}>0$
  \begin{align}
 W_{\rm{mp}}( \overline{ \mathbf{U}})=0\quad \Rightarrow\quad \overline{\mathbf{R}}^T\mathbf{F}-\id=0 \quad \Rightarrow\quad \overline{\mathbf{R}}=\mathbf{F}={\rm D}\varphi
 \end{align}
 and by taking the $\Curl$-operator on both sides we obtain
 \begin{align}
 W_{\rm{mp}}(\overline{ \mathbf{U}})=0\quad \Rightarrow\quad \Curl\, \overline{\mathbf{R}}=0.
 \end{align}
 Since $\overline{\mathbf{R}}\in {\rm SO}(3)$, this implies that $\overline{\mathbf{R}}$ is constant \cite{Neff_curl06} and yields 
  \begin{align}
 W_{\rm{mp}}( \overline{ \mathbf{U}})=0\quad \Rightarrow\quad \boldsymbol{\alpha}=0 \qquad (\boldsymbol{\mathfrak{K}}=0).
 \end{align}
 
 The same redundancy is true for the linear Cosserat model (in which $\mu_{\rm c}>0$ a priori). In this case we note that
 \begin{align}
\mathbf{e}=0 \quad \Rightarrow\quad {\rm D} u=\mathbf{A} \quad \Rightarrow\quad 0=\Curl\, \mathbf{A}=\alpha \quad \Rightarrow\quad \mathbf{A}={\rm constant}\quad \Rightarrow\quad {\rm D}\axl(\mathbf{A})=0 \qquad   (\boldsymbol{\mathfrak{K}}=0).
 \end{align}

Note that the linear relaxed micromorphic model is non-redundant for $\mu_{\rm c}=0$, a choice which is permitted for well-posedness.

\renewcommand{\arraystretch}{1.1}
\begin{table}[h!]
	\centering
	\begin{tabular}{|l|c|c|c|c|c|c|c|c|}
		\hline
		\multicolumn{1}{|c|}{Material}  & $E$ [MPa] & $G$ [MPa] & $\nu$ [-] & $N^2$ [-] & $\ell_t$ [mm] & $\ell_b$ [mm] & $\Psi$ [-]
		\\
		\hline
		\begin{tabular}[c]{@{}l@{}}Human bone \\ @0.2mm \end{tabular}
		& 12000
		& 4000
		& 0.5
		& 0.5
		& 0.22
		& 0.45
		& 3/2
		\\ \hline
		\begin{tabular}[c]{@{}l@{}}Graphite @1.6mm\\ (H237)\end{tabular}
		& 4500
		& 2122.64
		& 0.06
		& 1
		& 1.6
		& 2.8
		& (3/2)\,$^*$
		\\ \hline
		\begin{tabular}[c]{@{}l@{}}{Foam} @1mm\\ (0.6 PS)\end{tabular}
		& 1.28
		& 0.6
		& 0.07
		& 0.09
		& 3.8
		& 5
		& 3/2
		\\ \hline
		\begin{tabular}[c]{@{}l@{}}Foam @0.18mm\\ (dense polyurethane)\end{tabular}
		& 300
		& 104
		& 0.4
		& 0.04
		& 0.62
		& 0.33
		& 3/2
		\\ \hline
		\begin{tabular}[c]{@{}l@{}}{Foam} @0.15mm\\ (dense syntactic)\end{tabular}
		& 2758
		& 1033
		& 0.34
		& 0.1
		& 0.065
		& 0.0325
		& 3/2
		\\ \hline
	\end{tabular}
	\caption{
		In accordance with Lakes  \cite{Lakes95b}  (Table 1 at the end), Foam @1mm (0.6 PS) according to Lakes \cite{Lakes83} (pp. 2576-2577), and Foam @0.15mm (dense syntactic) according to \cite{Lakes85b} (p. 60).
		For Foam @0.15mm (dense syntactic) we have used $\ell_b=0.325$ \cite{Lakes85b}, instead of  $\ell_b=0.33$ included in \cite[Table 1]{Lakes95b}, since otherwise the positive semi-definiteness is violated  as well as the identified condition in bending experiment for this material \cite[p. 60]{Lakes95b}, i.e., $\beta/\gamma=1$.
		$^*$ Note that for  Graphite the Cosserat couple modulus $\mu_{\rm c}\to \infty$, a constraint that reduces the Cosserat model to its particular case, i.e., the couple stress model. In this case, the moment part of the stress tensor is trace free, consistent with $\Psi=3/2$ ($a_3=0$).}
	\label{tab:exp_values_papers}
\end{table}
\renewcommand{\arraystretch}{1}

\renewcommand{\arraystretch}{1.1}

\renewcommand{\arraystretch}{1}
 
\section{Conclusions}\label{conclusion}

In the literature on linear isotropic Cosserat or micropolar solids  many different abbreviations and definitions are frequently encountered. For convenience of the reader, especially for interpreting the experimental results \cite{Lakes81,Lakes83,Lakes85,Lakes85b,Lakes87,Lakes94,Lakes95b,lakes2016physical}, based on the analysis presented in the present paper, we collect these technical constants in Table \ref{tabelC}. We mention that further numerical values are proposed in \cite{Gauthier75,taliercio2010torsion,izadi2021torsional}.

The Cosserat theory of isotropic elastic solid written in the dislocation format (our proposal) represents a direct particular case of both the relaxed micromorphic model and of the geometrically nonlinear isotropic Cosserat model which uses the second order dislocation density tensor $\boldsymbol{\alpha}\,=\,\overline{\mathbf{R}}^T\, \Curl \,\overline{\mathbf{R}}\in\mathbb{R}^{3\times 3}$ as curvature measure.

We  remark that 
$
0\leq N\leq 1
$
and the limit case $N=1$ ($\mu_{\rm c}\to\infty$) corresponds to the case of the couple stress theory \cite{c1,c2,c3,c4}.
The technical parameter $\Psi$ (polar ratio) is positive if the necessary and sufficient conditions  \eqref{pden} for positive definiteness of the internal energy density are satisfied. However $\Psi$ is  not positive if the conditions 
\eqref{d11} for existence of real planar waves in any direction are satisfied  and neither when the Legendre-Hadamard ellipticity condition \eqref{dse1} are satisfied, since both these constitutive conditions imply the positivity of $2\,\alpha_1+\alpha_3$ but they do not imply the positivity of $\alpha_1$.

Therefore, assuming the internal energy to be positive definite  (the same remains true if $\alpha_2\geq 0$) the following one-to-one relation between our constitutive parameters and the constitutive technical constants considered by Lakes hold
\begin{align}
\lambda_{\rm e}=&-\frac{G \,(E-2 \,G)}{E-3 \,G}, \qquad\qquad\mu_{\rm e}= G,\qquad \qquad\qquad\qquad\mu _c=\frac{G \,N^2}{1-N^2},\notag\\\alpha _1=& \frac{2 \,\ell_t^2}{L_{\rm c}^2},\qquad\qquad\qquad\qquad\ \ \alpha _2=\frac{8\, \ell_b^2-2\, \ell_t^2}{L_{\rm c}^2},\qquad\qquad\alpha _3= \frac{4 \,\ell_t^2 (1-\psi)}{\psi \, L_{\rm c}^2},\\
a_1=&\frac{2 \,\ell_t^2}{L_{\rm c}^2},\qquad\qquad\qquad\qquad\ \, \ a _2=\frac{8\, \ell_b^2-2\, \ell_t^2}{L_{\rm c}^2},\qquad\qquad {a_3} =\frac{4 \,\ell_t^2 (3-2\,\psi)}{\psi \, L_{\rm c}^2}.\notag
\end{align}

 \begin{table}[]
	\centering
\resizebox{17cm}{!}{	\begin{tabular}{|l|c|c|c|c|c|c|c|c|c|}
		\hline
		\multicolumn{1}{|c|}{Material}                                         & $\mu_{\rm e}$ [MPa] & $\lambda_{\rm e}$ [MPa] & $\mu_{\rm c}$ [MPa] & \begin{minipage}{3cm}\medskip\centering$\mu_{\rm e}\,L_{\rm c}^2 \, \alpha_1$ [N]\\$\ \ \ =\boldsymbol{\mu_{\rm e}\,L_{\rm c}^2 \, a_1}$ [N] \end{minipage}& \begin{minipage}{3cm}\medskip\centering$\mu_{\rm e}\,L_{\rm c}^2 \, \alpha_2$ [N]\\$\ \ \ =\boldsymbol{\mu_{\rm e}\,L_{\rm c}^2 \, a_2}$ [N] \end{minipage}& $\mu_{\rm e}\,L_{\rm c}^2 \, \alpha_3$ [N]& $\boldsymbol{\mu_{\rm e}\,L_{\rm c}^2 \, a_3}$ [N]
		\\
		\hline
		\begin{tabular}[c]{@{}l@{}}Human bone \\ @0.2mm \end{tabular}
		& 4000
		& $\infty$ ($\nu=1/2$)
		& 4000
		& 387.2
		& 6092.8
		& -258.133&0
		\\ \hline
		\begin{tabular}[c]{@{}l@{}}Graphite @1.6mm\\ (H237)\end{tabular}
		& 2122.64
		& 289.451
		& 
		$\infty$ 
		& 10867.9
		& 122264
		& (-16301.85)\,$^*$&(0)\,$^*$
		\\ \hline
		\begin{tabular}[c]{@{}l@{}}{Foam} @1mm\\ (0.6 PS)\end{tabular}
		& 0.6
		& 0.0923077
		& 0.0593407
		& 17.328
		& 102.672
		& -11.552&0
		\\ \hline
		\begin{tabular}[c]{@{}l@{}}Foam @0.18mm\\ (dense polyurethane)\end{tabular}
		& 104
		& 797.333
		& 4.33333
		& 79.9552
		& 10.6496
		& -53.3035&0
		\\ \hline
		\begin{tabular}[c]{@{}l@{}}{Foam} @0.15mm\\ (dense syntactic)\end{tabular}
		& 1033
		& 2096.29
		& 114.778
		& 8.72885
		& 0
		& -5.81923&0
		\\ \hline
	\end{tabular}}


	\caption{The numerical experimental values of the new constitutive parameters according to Lakes  \cite{Lakes95b}. $^*$It is possible to evaluate $\alpha_3$ since $\mu_{\rm c}\to\infty$ implies that the moment part of the stress tensor is trace free, i.e., $\Psi=3/2$. The last row corresponds to conformal curvature. }
	\label{tab:exper_values_ours}
\end{table}

In the conformal curvature case \cite{Neff_JeongMMS08,Jeong_Neff_ZAMM08} \begin{align}\label{confa}
\alpha_2=0\,\Leftrightarrow\,a_2=0\,\Leftrightarrow\,\ell_t=2\,\ell_b\,\Leftrightarrow\,\beta=\gamma \qquad \text{\bf and}\qquad  2\,\alpha_1+3\,\alpha_3=0\,\Leftrightarrow \,a_3=0\,\Leftrightarrow \,\Psi=\frac{3}{2},
\end{align}
 the internal energy density is
\begin{align}
W
=&\,\mu_{\rm e} \,\lVert\dev   \, \sym\,\mathbf{e}\rVert ^{2}+\mu_{\rm c}\,\lVert \skw\,\mathbf{e}\rVert ^{2}+\frac{2\mu_{\rm e} +3\,\lambda_{\rm e}  }{6}\left[\mathrm{tr} \left(\mathbf{e}\right)\right]^{2} +\frac{\mu_{\rm e} L_{\rm c}^2}{2}\,{a_1}\|\dev  \,\sym \,\Curl\, \mathbf{A}\|^2\notag\\
=&\,\mu_{\rm e} \,\lVert\dev   \, \sym\,\mathbf{e}\rVert ^{2}+\mu_{\rm c}\,\lVert \skw\,\mathbf{e}\rVert ^{2}+\frac{2\mu_{\rm e} +3\,\lambda_{\rm e}  }{6}\left[\mathrm{tr} \left(\mathbf{e}\right)\right]^{2} +\frac{\mu_{\rm e} L_{\rm c}^2}{2}\,{\alpha_1}\|\dev  \,\sym \,\Curl\, \mathbf{A}\|^2\\
=&\, \mu_{\rm e} \,\lVert\dev   \, \sym\,\mathbf{e}\rVert ^{2}+\mu_{\rm c}\,\lVert \skw\,\mathbf{e}\rVert ^{2}+\frac{2\mu_{\rm e} +3\,\lambda_{\rm e} }{6}\left[\mathrm{tr} \left(\mathbf{e}\right)\right]^{2} +\frac{\mu_{\rm e} L_{\rm c}^2}{2}\, {\alpha_1}\,\lVert\dev   \,\sym\,\mathbf{\mathfrak{K}}\rVert ^{2} \, .
\notag
\end{align}

Therefore, the conformal curvature case \cite{neff2010stable} is characterised by the technical constants\footnote{These two conditions are independent.} 
\begin{align}
\ell_t=2\,\ell_b \qquad(\beta=\gamma) \qquad \text{and} \qquad \qquad \Psi=\frac{3}{2}.
\end{align} 
This is the case of  Foam @0.15mm (dense syntactic) and nearly satisfied for Foam @0.18mm. We also note that $\Psi= \frac{3}{2}$ corresponds to the curvature energy $\frac{\mu_{\rm e} L_{\rm c}^2}{2}\,{\alpha_1}\|\dev  \,\sym \,\Curl\, \mathbf{A}\|^2$  and is sufficient for bounded stiffness in torsion \cite{Neff_JeongMMS08}, 
while $\beta=\gamma$ ($a_2=0$) is necessary for bounded stiffness in bending.
Both limit cases violate uniform positivity of the curvature but are allowed by the weaker requirements in  \eqref{posCoss1}.

Under uniform positive definiteness of the energy, the admitted range for $\Psi$ is (see Table \ref{tabelC})
\begin{align}
0<\Psi<\frac{3}{2} \qquad \qquad (\text{together with}\quad \beta-\gamma>0).
\end{align}
According to our weaker requirements \eqref{ew} the admitted range for $\Psi$ is
\begin{align}
0<\Psi\leq\frac{3}{2}\quad  (a_3=0)\qquad \qquad (\text{and}\quad \beta-\gamma=0,\  \text{too}).
\end{align}
The latter limit value $\Psi=\frac{3}{2}$ ($a_3=0$) has been consistently taken in the identification by Lakes, see Tables \ref{tab:exp_values_papers} and \ref{tab:exper_values_ours}.

If only one  of the two conditions \eqref{confa} characterising the conformal curvature case  is satisfied we have: for  $\alpha_2=0\,\Leftrightarrow\,a_2=0\,\Leftrightarrow\,\ell_t=2\,\ell_b\,\Leftrightarrow\,\beta=\gamma$   the internal energy density is
\begin{align*}
W=& \,\mu_{\rm e} \,\lVert\dev   \, \sym\,\mathbf{e}\rVert ^{2}+\mu_{\rm c}\,\lVert \skw\,\mathbf{e}\rVert ^{2}+\frac{2\mu_{\rm e} +3\,\lambda_{\rm e}  }{6}\left[\mathrm{tr} \left(\mathbf{e}\right)\right]^{2}+\frac{\mu_{\rm e} L_{\rm c}^2}{2}\left[{a_1}\|\dev  \,\sym \,\Curl\, \mathbf{A}\|^2 + \frac{a_3}{3}\, \tr(\Curl\, \mathbf{A})^2\right]\notag\\
=&\, \mu_{\rm e} \,\lVert\dev   \, \sym\,\mathbf{e}\rVert ^{2}+\mu_{\rm c}\,\lVert \skw\,\mathbf{e}\rVert ^{2}+\frac{2\mu_{\rm e} +3\,\lambda_{\rm e} }{6}\left[\mathrm{tr} \left(\mathbf{e}\right)\right]^{2} +\frac{\mu_{\rm e} L_{\rm c}^2}{2}\,\left[\ {\alpha_1}\,\lVert\dev   \,\sym\,\mathbf{\mathfrak{K}}\rVert ^{2}+\frac{2\,\alpha_1+3\,\alpha_3}{6}\left[\mathrm{tr} \left(\mathbf{\mathfrak{K}}\right)\right]^{2}\right],\notag
\end{align*}
while for  $2\,\alpha_1+3\,\alpha_3=0\,\Leftrightarrow \,a_3=0\,\Leftrightarrow \,\Psi=\frac{3}{2}$ the internal energy density is
\begin{align*}
W=& \,\mu_{\rm e} \,\lVert\dev   \, \sym\,\mathbf{e}\rVert ^{2}+\mu_{\rm c}\,\lVert \skw\,\mathbf{e}\rVert ^{2}+\frac{2\mu_{\rm e} +3\,\lambda_{\rm e}  }{6}\left[\mathrm{tr} \left(\mathbf{e}\right)\right]^{2} +\frac{\mu_{\rm e} L_{\rm c}^2}{2}\left[{a_1}\|\dev  \,\sym \,\Curl\, \mathbf{A}\|^2 +{a_2}\| \skw \,\Curl\, \mathbf{A}\|\right]\notag\\
=&\, \mu_{\rm e} \,\lVert\dev   \, \sym\,\mathbf{e}\rVert ^{2}+\mu_{\rm c}\,\lVert \skw\,\mathbf{e}\rVert ^{2}+\frac{2\mu_{\rm e} +3\,\lambda_{\rm e} }{6}\left[\mathrm{tr} \left(\mathbf{e}\right)\right]^{2} +\frac{\mu_{\rm e} L_{\rm c}^2}{2}\,\left[\ {\alpha_1}\,\lVert\dev   \,\sym\,\mathbf{\mathfrak{K}}\rVert ^{2}+{\alpha_2}\,\lVert \skw\,\mathbf{\mathfrak{K}}\rVert ^{2}\right].\notag
\end{align*}

Using these direct identification, in Table \ref{tab:exper_values_ours} we provide  the numerical values of the material parameters considered in our formulation of the Cosserat theory of isotropic elastic solid, according to the experimental results obtained in  \cite{Lakes95b} (see Table \ref{tab:exp_values_papers}).

\paragraph*{{Acknowledgements.}} The authors are indebted to Roderic Lakes (Wisconsin Distinguished Professor, University of Wisconsin-Madison) for the motivation to provide this comprehensive comparison. The  work of I.D. Ghiba  was supported by a grant of the Romanian Ministry of Research
and Innovation, CNCS--UEFISCDI, Project no.
PN-III-P1-1.1-TE-2019-0348, Contract No. TE 8/2020, within PNCDI III. 
 Angela Madeo and Gianluca Rizzi acknowledges support from the European Commission through the funding of the ERC Consolidator Grant META-LEGO, N° 101001759.
Angela Madeo and Gianluca Rizzi acknowledge funding from the French Research Agency ANR, “METASMART” (ANR-17CE08-0006).
Angela Madeo thanks IUF (Institut Universitaire de France) for its support.
Patrizio Neff acknowledges support in the framework of the DFG-Priority Programme 2256 ``Variational Methods for Predicting Complex Phenomena in Engineering Structures and Materials'', Neff 902/10-1, Project-No. 440935806.




	\bigskip

\bibliographystyle{plain} 

\addcontentsline{toc}{section}{References}
\begin{footnotesize}

\end{footnotesize}
\end{document}